\documentclass[prl,twocolumn,amsmath,nobibnotes,nofootinbib,superscriptaddress,preprintnumbers,hyperref]{revtex4-1}

\usepackage{bm} \usepackage{graphicx} \usepackage{color}
\usepackage{epstopdf} \usepackage{amssymb}
\def\ba{\begin{eqnarray}}
\def\ea{\end{eqnarray}}
\def\be{\begin{equation}}
\def\ee{\end{equation}}
\def\({\left(}
\def\){\right)}
\def\[{\left[}
\def\]{\right]}
\def\<{\left<}
\def\>{\right>}

\newcommand{\ellmax} {\ell_{\rm max}}
\newcommand{\ellmaxrecon} {\ell_{\rm max,\,rec}}
\newcommand{\cl} {C_{\ell}}
\newcommand{\clsmooth} {\bar{C}_{\ell}}
\newcommand{\clqml} {\hat{C}_{\ell}}
\newcommand{\alm} {a_{\ell m}}
\newcommand{\almest} {\hat{a}_{\ell m}}

\newcommand{\epslm} {\epsilon_{\ell m}}

\newcommand{\almvec} {\mathbf{a}}
\newcommand{\almestvec} {\mathbf{\hat{a}}}

\newcommand{\epslmvec} {\bm{\epsilon}}

\newcommand{\wmat} {\mathbf{W}}
\newcommand{\ymat} {\mathbf{Y}}
\newcommand{\cmat} {\mathbf{C}}
\newcommand{\smat} {\mathbf{S}}
\newcommand{\rmat} {\mathbf{R}}
\newcommand{\nmat} {\mathbf{N}}
\newcommand{\datavec} {\mathbf{x}}
\newcommand{\pixnoisevec} {\mathbf{n}}
\newcommand{\posvec} {\mathbf{\hat{r}}}

\newcommand{\fsky} {f_{\rm sky}}
\newcommand{\nalm} {n_{a_{\ell m}}}
\newcommand{\npix} {n_{\rm pix}}
\newcommand{\nside} {N_{\rm side}}
\newcommand{\zlm} {z_{\ell m}}
\newcommand{\ztot} {Z}

\newcommand{\biasvec} {\mathbf{b}}

\newcommand{\order} {\mathcal{O}}
\newcommand{\muk} {\mu\mathrm{K}}

\begin{document}

\title{Avoiding bias in reconstructing the largest observable scales from partial-sky data}
\date{\today}

\author{Stephen M. Feeney}
\email{stephen.feeney.09@ucl.ac.uk}
\affiliation{Department of Physics and Astronomy, University College London, London WC1E 6BT, U.K.}
\author{Hiranya V. Peiris}
\email{h.peiris@ucl.ac.uk}
\affiliation{Department of Physics and Astronomy, University College London, London WC1E 6BT, U.K.}
\author{Andrew Pontzen}
\email{apontzen@ast.cam.ac.uk}
\affiliation{Institute of Astronomy and Kavli Institute for Cosmology, University of Cambridge, Cambridge CB3 0HA, U.K.}

\begin{abstract}
Obscuration due to Galactic emission complicates the extraction of information from cosmological surveys, and requires some combination of the (typically imperfect) modeling and subtraction of foregrounds, or the removal of part of the sky. This particularly affects the extraction of information from the largest observable scales. Maximum-likelihood estimators for reconstructing the full-sky spherical harmonic coefficients from partial-sky maps have recently been shown to be susceptible to contamination from within the sky cut, arising due to the necessity to band-limit the data by smoothing prior to reconstruction. Using the WMAP 7-year data, we investigate modified implementations of such estimators which are robust to the leakage of contaminants from within masked regions. We provide a measure, based on the expected amplitude of residual foregrounds, for selecting the most appropriate estimator for the task at hand. We explain why the related quadratic maximum-likelihood estimator of the angular power spectrum does not suffer from smoothing-induced bias.
\end{abstract}

\preprint{}

\maketitle

\section{Introduction}

It is unavoidable that we observe the Universe through the galaxy we inhabit. The foreground contamination injected by the Milky Way into full-sky cosmological data-sets must be modeled and removed, or the regions most conspicuously contaminated must be excised. Where no precise model of the foregrounds is available, cutting the sky is the most robust option, with the regrettable consequence that part of the signal is discarded along with the contamination. This includes information on the largest scales, which are valuable for a variety of reasons, including measurement of the integrated Sachs-Wolfe effect~\cite{1967ApJ...147...73S} and constraining primordial non-Gaussianity using tracers of large-scale structure~\cite{Slosar:2008hx}.

It is impossible to uniquely recover the cosmological signal discarded in the sky cut. However, by writing down the likelihood for the region of the sky in which one trusts the data, it is possible to reconstruct an estimate of the signal at large scales which maximizes the likelihood of the residual noise~\cite{deOliveiraCosta:2006zj}. An alternative reconstruction scheme maximizes the {\em posterior} probability~\cite{Wiener:1964, Bunn:1994xn, Zaroubi:1994mx, Tegmark:1996qs, Bielewicz:2004en} of measuring the underlying cosmological signal given the available data  and a prior theoretical expectation on the signal.

The reconstructions estimate the large-scale (low-$\ell$) spherical harmonic coefficients, $\alm$, by treating the signal at small scales as noise and only considering data external to the sky cut. If, as with the cosmic microwave background (CMB), the field to be reconstructed is not band-limited, the proliferation of small-scale signal makes the reconstruction noisy to the point of being useless. Input maps are therefore smoothed -- necessarily prior to cutting the sky 
 -- to truncate the signal and remove sources of confusion below a chosen angular scale~\cite{Efstathiou:2009di}. However, smoothing leaks contamination from the masked region into the trusted data~\cite{Aurich:2010gw,Copi:2011pe}, and the reconstructed spherical harmonic coefficients, $\almest$, are biased. In this work we explore the causes and expected magnitudes of this bias, and discuss how it can be mitigated.

\section{Maximum-Likelihood Reconstruction}

We begin with a description of the standard implementation of maximum-likelihood CMB $\alm$ reconstruction. The first step of the reconstruction process is to band-limit the temperature field by smoothing, typically with a Gaussian kernel of width $10^{\circ}$ FWHM. As this removes information on the smallest scales, the map resolution can be downgraded to reduce computation time. The $\almest$s in the range $2 \le \ell \le \ellmaxrecon$ (represented for ease as the $\nalm$-element vector $\almestvec$, where $\nalm = (\ellmaxrecon - 1)(\ellmaxrecon+3)$) are then reconstructed from the $\npix$ unmasked pixel temperatures, $\datavec$, using~\cite{deOliveiraCosta:2006zj}
\begin{equation}
\label{eq:alm_recon}
\almestvec = \wmat \datavec .
\end{equation}
The reconstructed spherical harmonic coefficients maximize the likelihood of the residual noise, given the available data, if the reconstruction matrix, $\wmat$, is
\begin{equation}
\label{eq:ml_recon_mat}
\wmat = [ \ymat^t \cmat^{-1} \ymat ]^{-1} \ymat^t \cmat^{-1}.
\end{equation}
Here, $\ymat$ are the spherical harmonics calculated at each unmasked pixel\footnote{Without loss of generality, the reconstruction matrix in this work is formed from the {\em real} spherical harmonics.}, and $\cmat$ is the pixel-space noise covariance matrix
\begin{equation}
\label{eq:recon_cmat}
C_{ij} = R_{ij} + \sum_{\ell = \ellmaxrecon + 1}^{\ellmax} \frac{2\ell + 1}{4\pi} \clsmooth P_{\ell}(\posvec_i \cdot \posvec_j),
\end{equation}
where $\rmat$ is uncorrelated, low-amplitude regularizing noise added to prevent $\cmat$ from becoming singular, $\clsmooth$ is the smoothed theory CMB angular power spectrum, and $P_{\ell}$ are the Legendre polynomials at unmasked pixels $i, j$. The sum over the multipoles $\ellmaxrecon < \ell \le \ellmax$ ensures that the small-angular-scale CMB multipoles we do not wish to reconstruct are treated as noise. As stated above the CMB power must be artificially truncated to restrict the number of ambiguous modes accessible to the reconstruction. The smoothing kernel is deconvolved from the $\almest$s after reconstruction by dividing the $\almest$s by the kernel's spherical harmonic transform. 

If a foreground signal $b_i$ is now introduced, so that $\datavec=\ymat \almvec + \biasvec + \pixnoisevec$ and $\almvec$ is the CMB signal uncorrelated with $\biasvec$ and the noise $\pixnoisevec$, the mean and variance of the reconstruction error $\epslm = \almest - \alm$ are
\begin{equation}
\label{eq:recon_err_ml_mean}
\langle \epslmvec \rangle = \wmat \biasvec
\end{equation}
and
\begin{equation}
\label{eq:recon_err_ml_var}
\langle \epslmvec \epslmvec^t \rangle - \langle \epslmvec \rangle \langle \epslmvec^{t} \rangle = \wmat \cmat \wmat^t,
\end{equation}
respectively.

Throughout this work, we reconstruct the spherical harmonic coefficients up to $\ellmaxrecon = 10$. The noise covariance matrix includes CMB power in the range $\ellmaxrecon < \ell \le \ellmax = 32$ unless explicitly stated; this value is chosen such that modes with $\ell > \ellmax$ are suppressed to $\order$(few $\%$) by the smoothing. The WMAP 5-year best-fit $\cl$s~\cite{Nolta:2008ih} are chosen for the theory CMB angular power spectrum\footnote{Our results are not sensitive to the small differences between different WMAP releases in the best-fit cosmology.}. Input maps are smoothed at HEALPix~\cite{Gorski:2004by} resolution $\nside = 512$ before being downgraded to $\nside = 16$ to retain the information required for the reconstruction while minimizing the number of pixels included in the noise covariance matrix. Diagonal regularizing noise $\rmat$ is added at the level of $2\,\muk^2$ to allow the inversion of the noise covariance matrix despite the presence of some null modes (which are irrelevant to the reconstruction).

\section{Smoothing-Induced Bias}

This section outlines how a bias arises from smoothing-induced contamination of the unmasked pixels \cite{Aurich:2010gw,Copi:2011pe}. For clarity, we illustrate the smoothing-induced bias in the maximum-likelihood reconstruction (and, later, our proposed solutions) with results plotted in both harmonic and pixel space.

It is beyond the scope of this work to estimate accurate foreground residuals resulting from different component separation methods; instead, we choose some residuals for the purpose of illustration. Following Ref.~\cite{Pontzen:2010uq}, the residual foregrounds are taken to be $1\%$ of the difference between the WMAP 7-year Internal Linear Combination (ILC)~\cite{Gold:2010fm} and V-band~\cite{Jarosik:2010iu} temperature maps. The resulting map is {\em indicative} of the extent and amplitude of the residual foregrounds in the WMAP 7-year ILC, the data-set for which the smoothing-induced bias was first described~\cite{Efstathiou:2009di,Aurich:2010gw,Copi:2011pe}. It is important to note that a considerably higher level of contamination is present in the foreground-reduced maps for each individual WMAP frequency band provided by the WMAP team: contaminants of $\sim 50$ times those used here are visible in these maps. The residual foregrounds are restricted to the pixels within the sky cut, which in this work masks only the Galaxy and not individual point-sources. These degree-scale point-source cuts are, unsurprisingly, not found to significantly bias the large-scale reconstructed $\almest$s, and so for clarity Galaxy-only contaminants and masks are considered.

\begin{figure}[tb]
\includegraphics[width=8.5 cm]{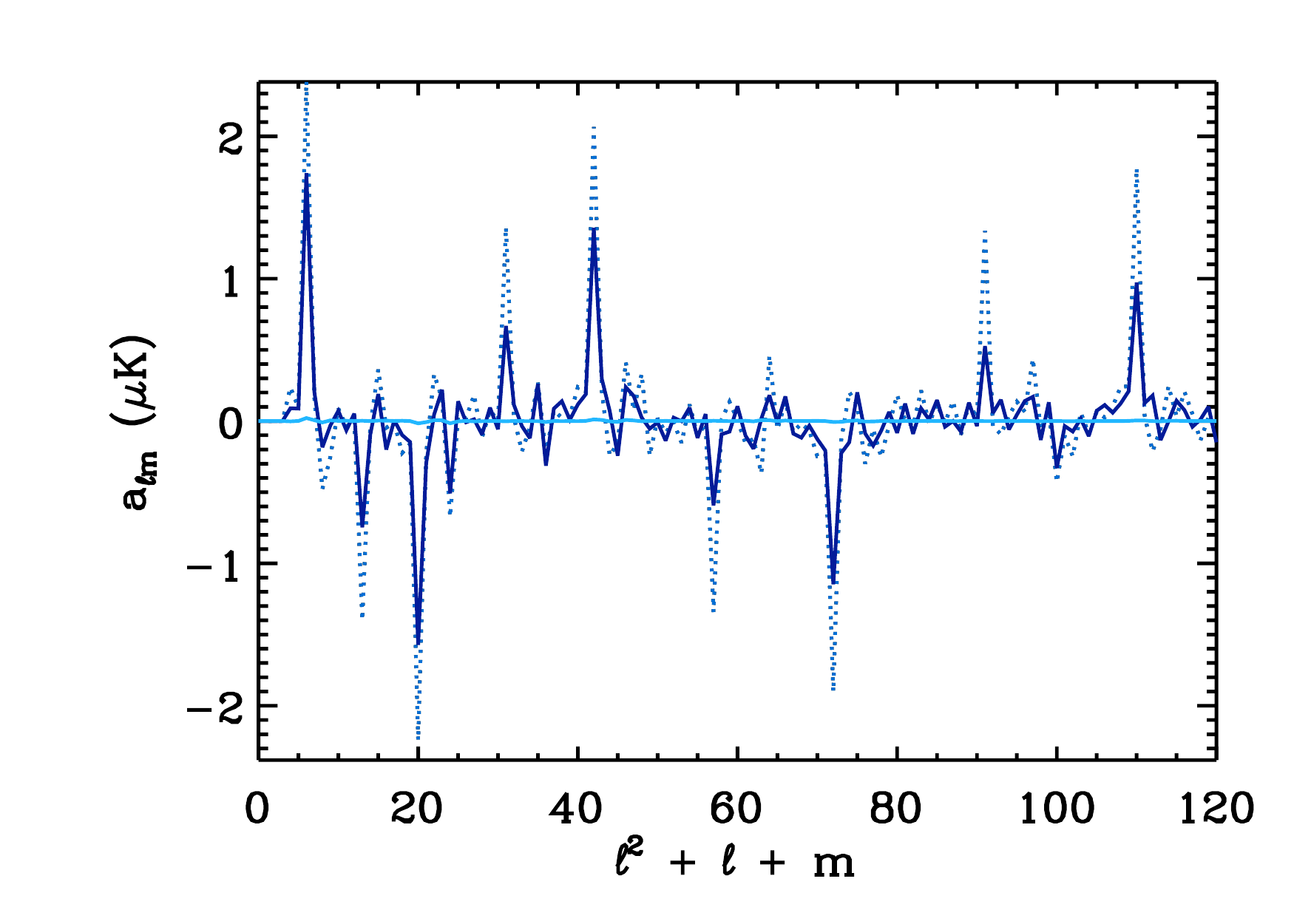}
\caption{The spherical harmonic coefficients of the simulated foreground residuals, calculated using the Gaussian ML reconstruction (deep-blue solid line), the full-sky data (dotted line) and the $10^{\circ}$ Top-Hat ML reconstruction (light-blue solid line on $x$-axis). It is clear that the Gaussian ML reconstruction leaks around half of the information from within the sky cut; this can be counteracted by smoothing with a top-hat kernel and using an extended mask. The HEALPix index $\ell ^ 2 + \ell + m$ maps each $\ell$, $m$ combination to a unique index into the array of $\alm$s.}
\label{fig:alm_est_bias}
\end{figure}

The addition of simulated residuals and the spherical harmonic transform are linear, so the smoothing-induced bias is given by the reconstructed $\almest$s of the simulated residual map. We take the ``standard'' $10^{\circ}$-FWHM-Gaussian-smoothed maximum-likelihood reconstruction, using data outside the Galaxy-only part of the WMAP 7-year KQ85 mask, as our fiducial maximum-likelihood estimator (hereafter the ``Gaussian ML'' $\almest$s). The Gaussian ML $\almest$s generated from the simulated residual foregrounds are plotted (deep-blue solid line) in Fig.~\ref{fig:alm_est_bias}, along with the full-sky $\alm$s (dotted line) for comparison. The smoothing-induced leakage from within the sky cut is clear to see: the reconstruction picks up about half of the power in the foreground residuals, even though the simulated residuals are entirely confined to the sky cut.

\begin{figure}[tb]
\includegraphics[width=4.25 cm]{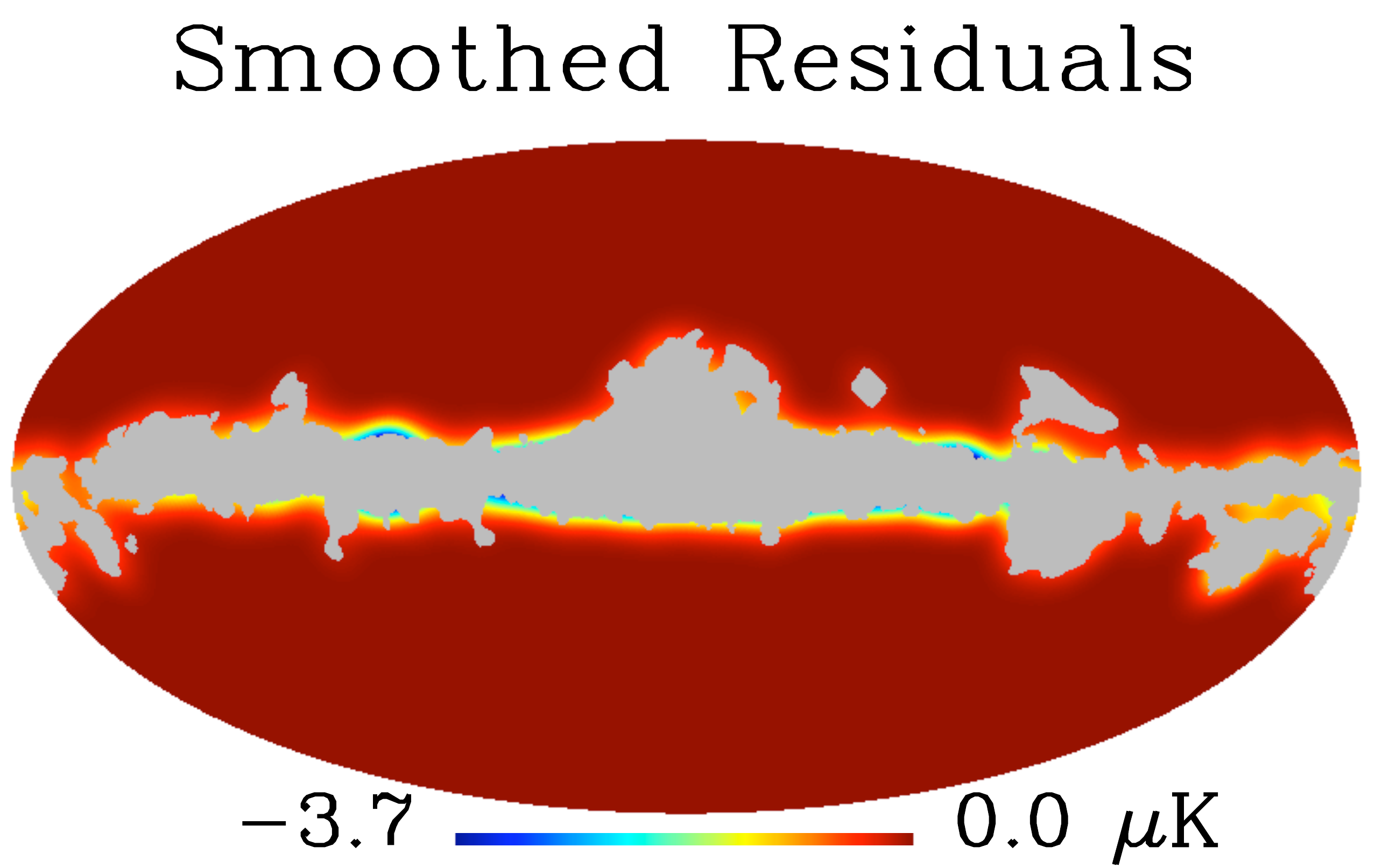}
\includegraphics[width=4.25 cm]{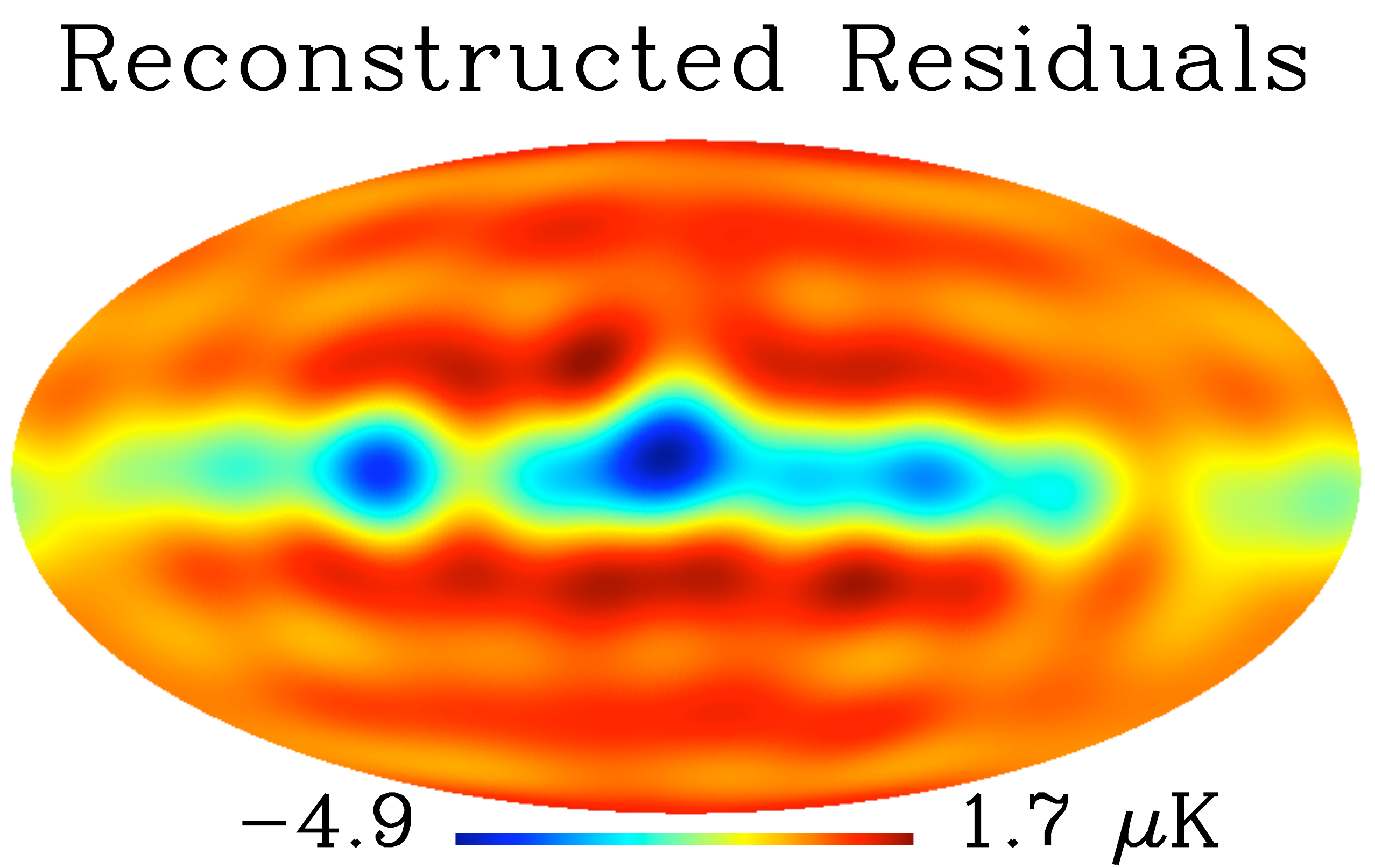}
\includegraphics[width=4.25 cm]{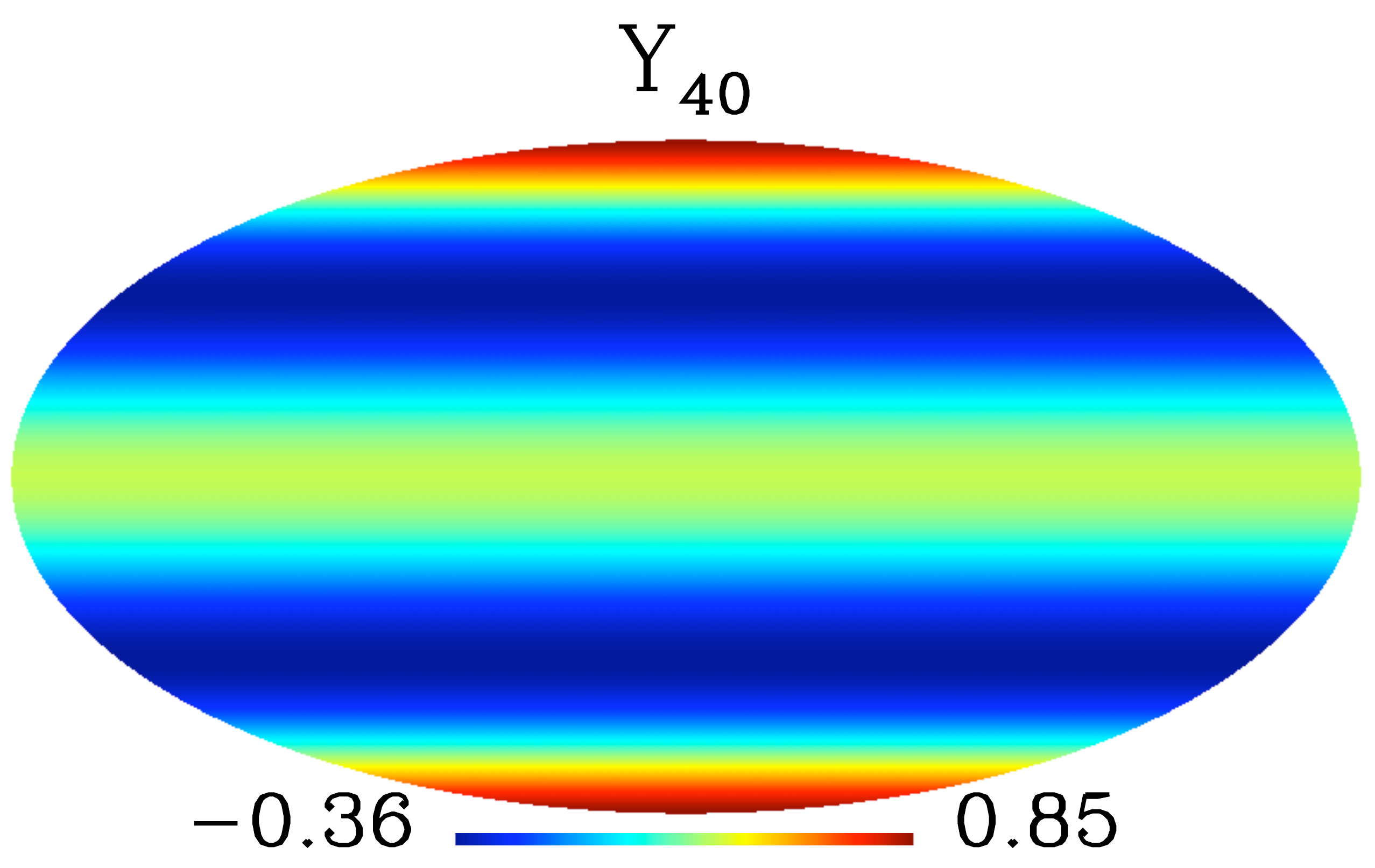}
\includegraphics[width=4.25 cm]{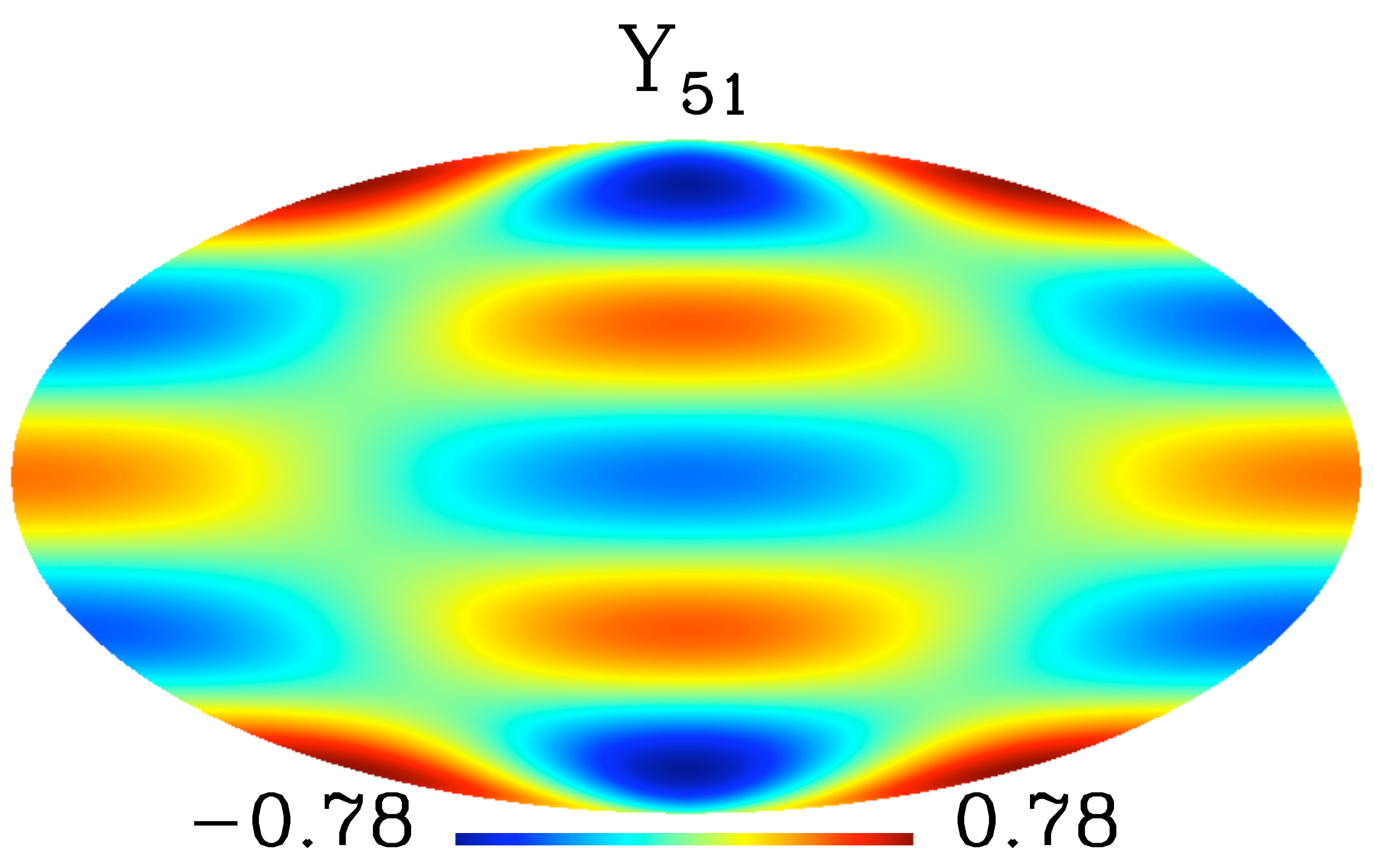}
\caption{Clockwise, from top left: simulated Galactic foregrounds, smoothed by a $10^{\circ}$-FWHM Gaussian and masked with the Galaxy-only KQ85 mask; the $\ell \le 10$ Gaussian ML $\almest$s reconstructed from the simulated Galactic foregrounds; the real spherical harmonics $Y_{4\,0}$ and $Y_{5\,1}$. The simulated foregrounds yield a negative bias in the $Y_{4\, 0}$ mode, and a positive bias in the $Y_{5\, 1}$ mode.}
\label{fig:res_ylm}
\end{figure}

The largest peaks in the bias affect the $\almest$s satisfying $\ell = 2n, m = 0$ for integer $n$ (in Fig.~\ref{fig:alm_est_bias}, HEALPix index $\ell^2 + \ell + m = \{6, 20, 42\ldots\}$)~\cite{deOliveiraCosta:2006zj}, and are positive for odd $n$, negative for even $n$. The pattern of these peaks can be explained by examining the reconstruction of the simulated Galactic residuals, plotted in Fig.~\ref{fig:res_ylm}, which are coldest along the Galactic plane. Smoothing these residuals reduces the pixel temperature values approximately symmetrically around the Galactic mask, and therefore pollutes the azimuthal modes which are also symmetric about the equator. The bias is positive for modes which have minima at the equator, and negative for those with maxima. There are also secondary peaks at $\ell = 2n + 1, m = 1$ (in Fig.~\ref{fig:alm_est_bias}, $\ell^2 + \ell + m = \{13, 31, 57\ldots\}$), which again are positive for odd $n$, negative for even $n$. These modes pick out the concentration of reconstructed foreground power in the Galactic centre.

\begin{figure}[tb]
\includegraphics[width=4.25 cm]{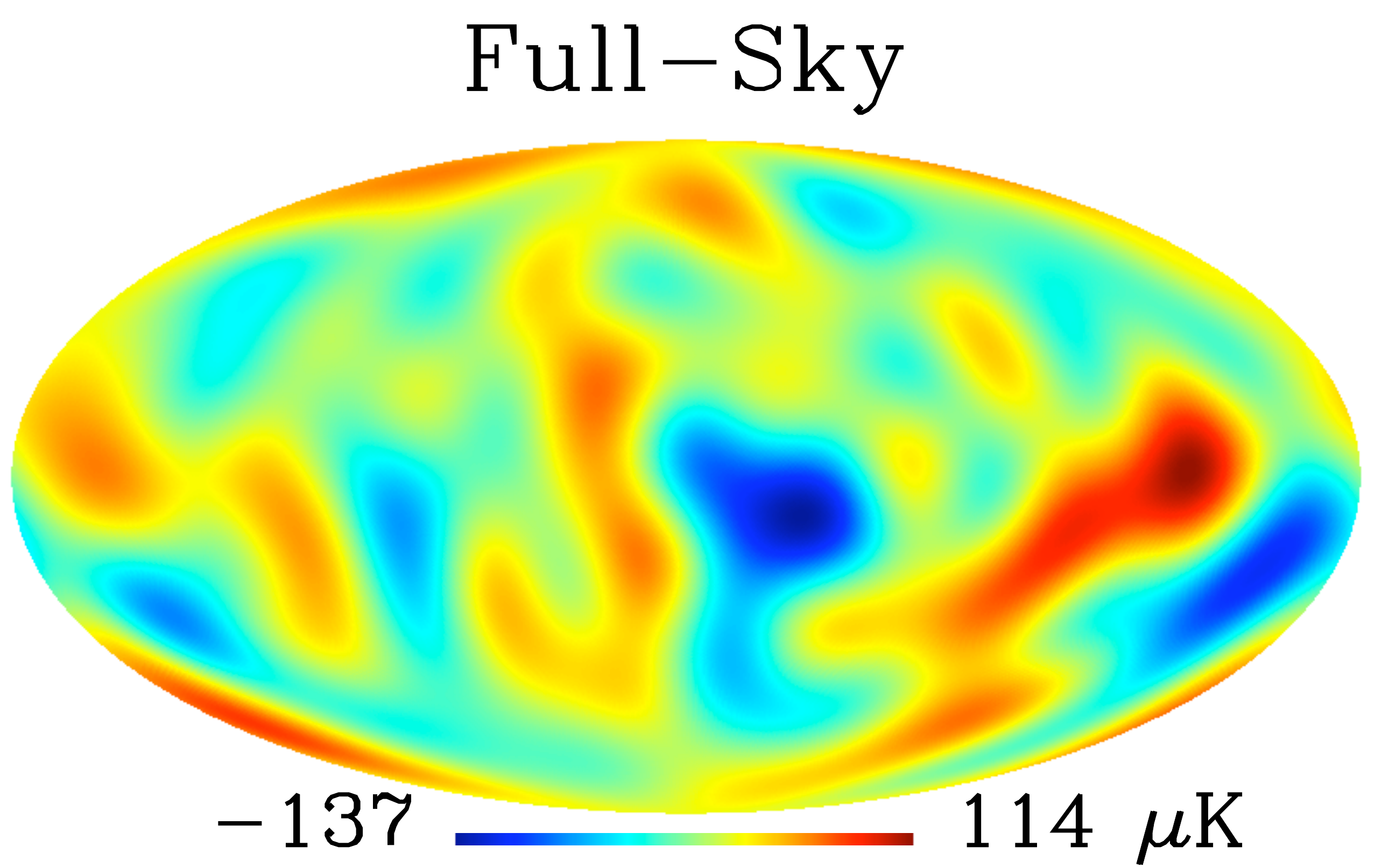}
\includegraphics[width=4.25 cm]{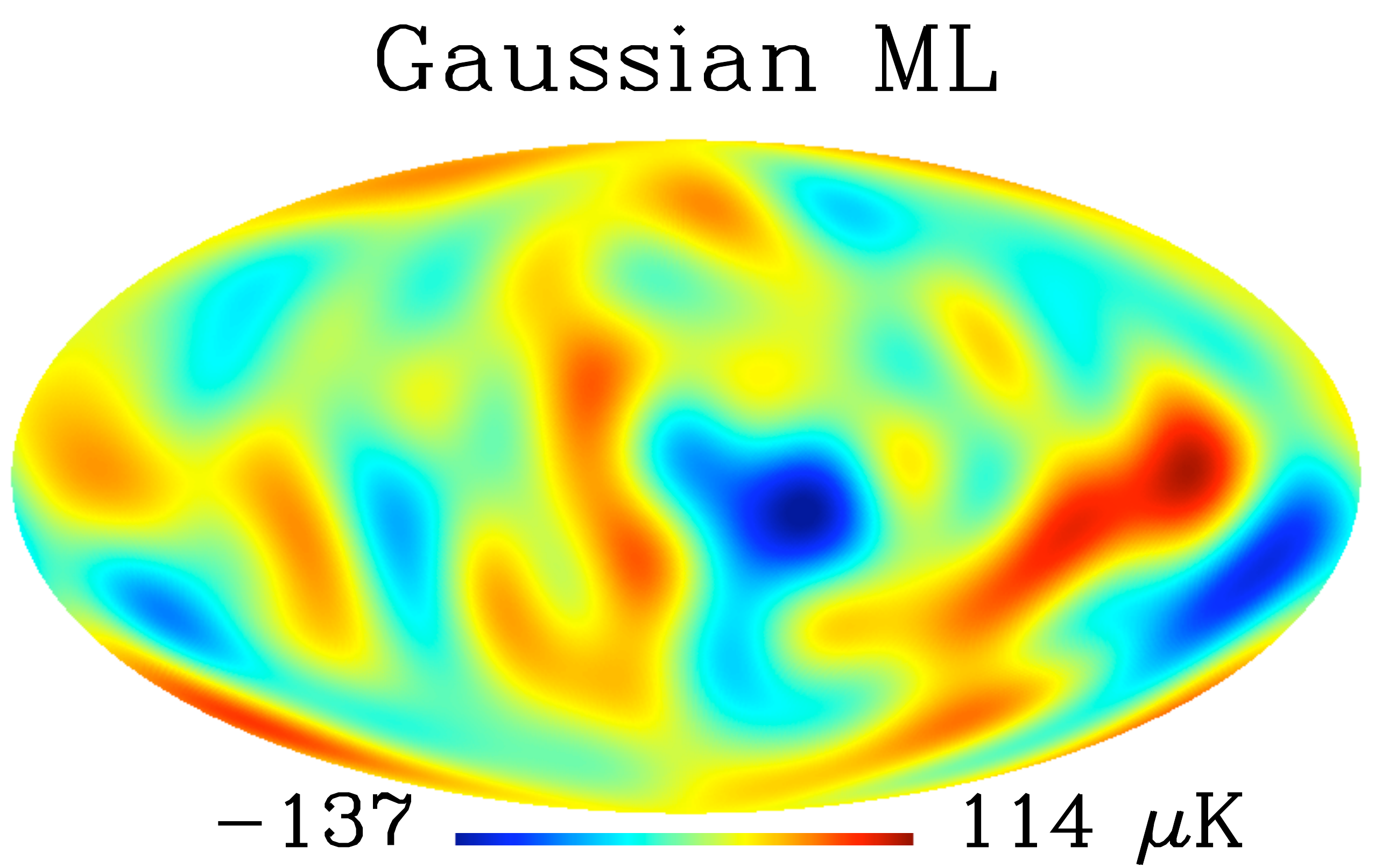}
\includegraphics[width=4.25 cm]{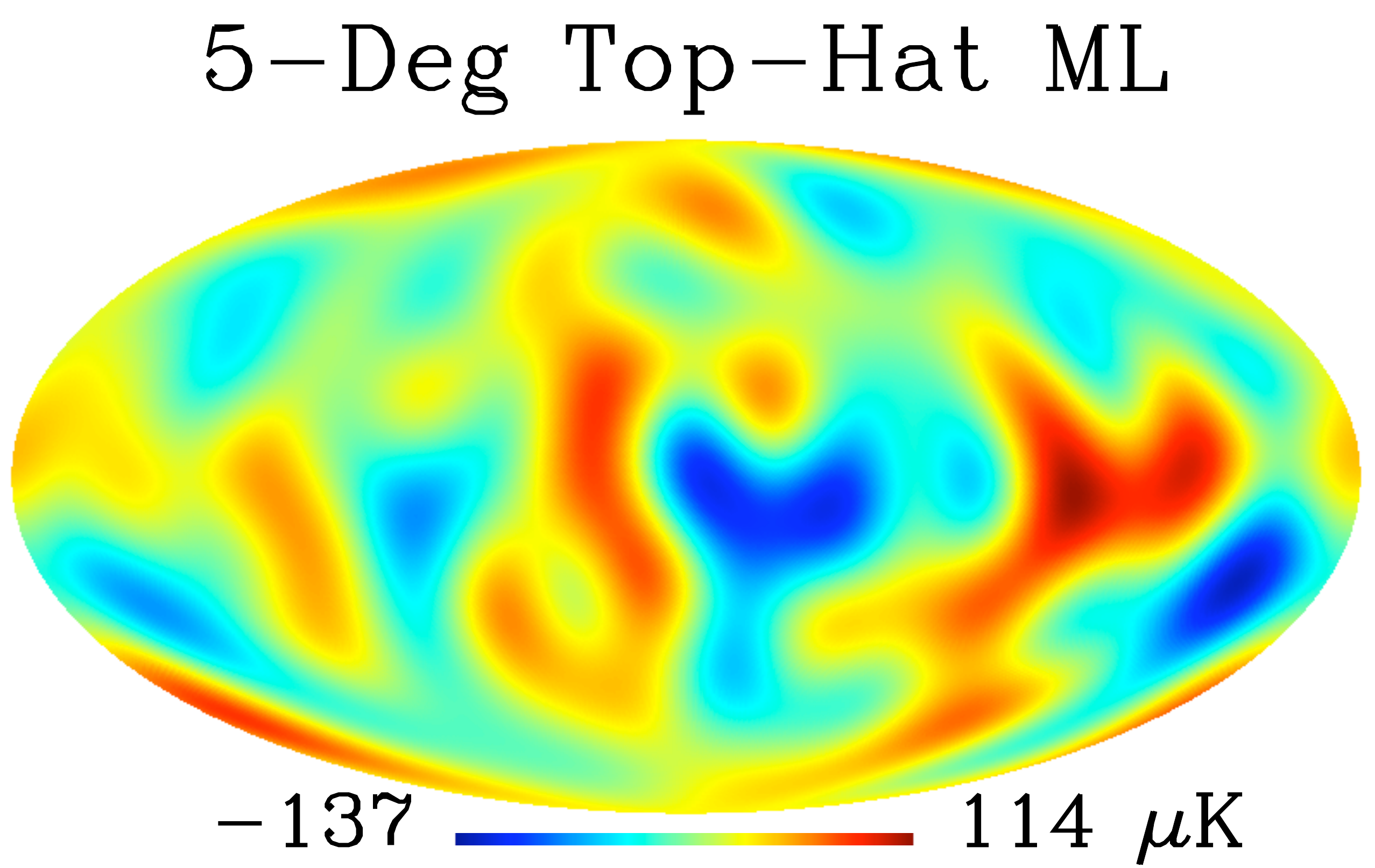}
\includegraphics[width=4.25 cm]{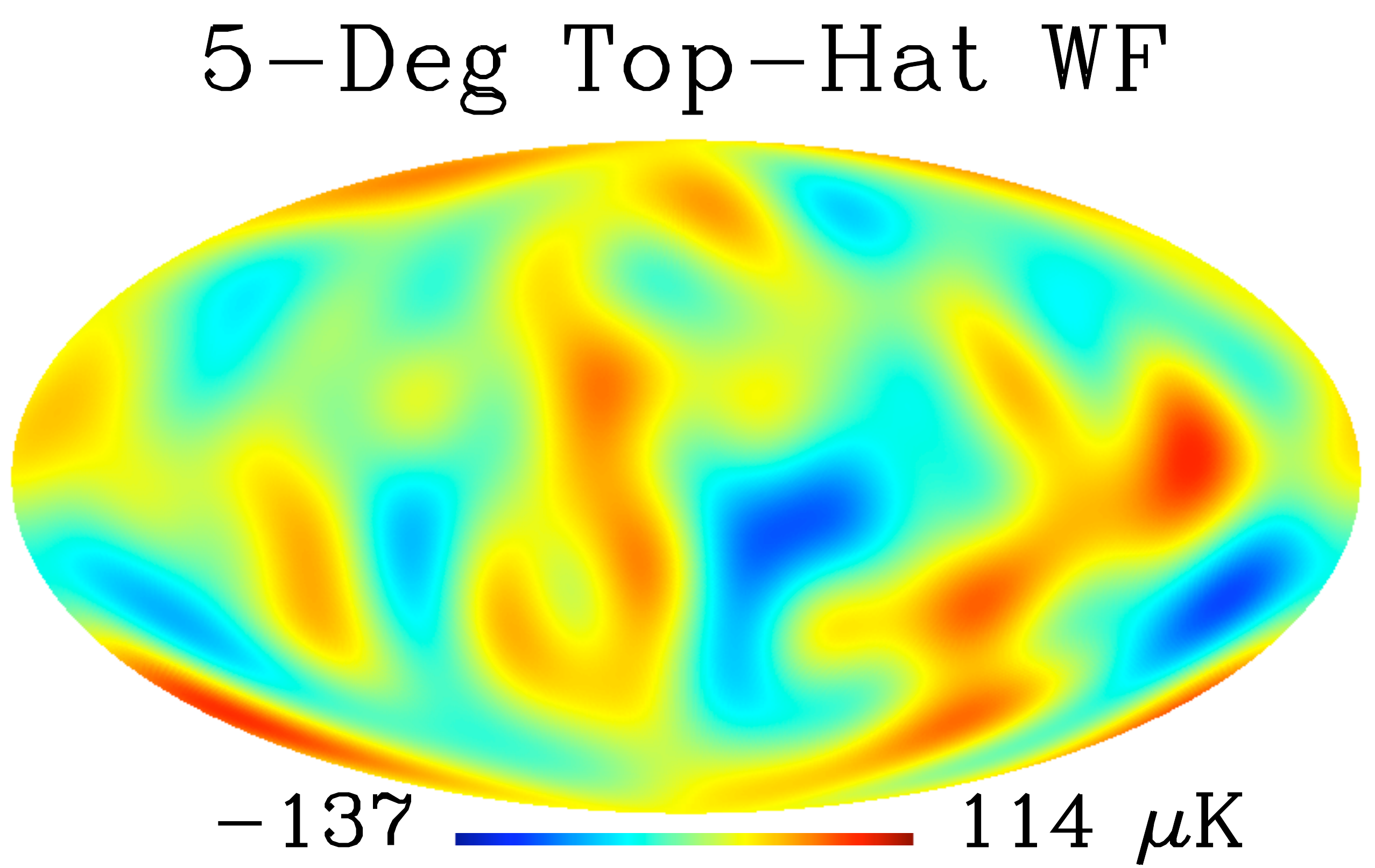}
\caption{The 7-year ILC $2 \le \ell \le 10$ modes, plotted using (clockwise from top-left) full-sky $\alm$s, Gaussian ML $\almest$s, $5^{\circ}$ Top-Hat ML $\almest$s, and Top-Hat WF $\almest$s.}
\label{fig:ilc_recon}
\end{figure}

The leakage of information from within the sky cut can also be demonstrated in pixel-space. Taking the WMAP 7-year  ILC map, the full-sky $\alm$s are extracted, and the Gaussian ML reconstruction is performed. The spherical harmonic coefficients recovered in each case are then used to reconstruct the input ILC map using only $2 \le \ell \le 10$, as plotted in Fig.~\ref{fig:ilc_recon}. The maps formed from the full-sky-$\alm$s (top-left) and reconstructed from Gaussian-ML-$\almest$s (top-right) are almost identical, even in the Galactic plane, confirming that the reconstruction has access to information well inside the sky cut.

\begin{figure}[tb]
\includegraphics[width=8.5 cm]{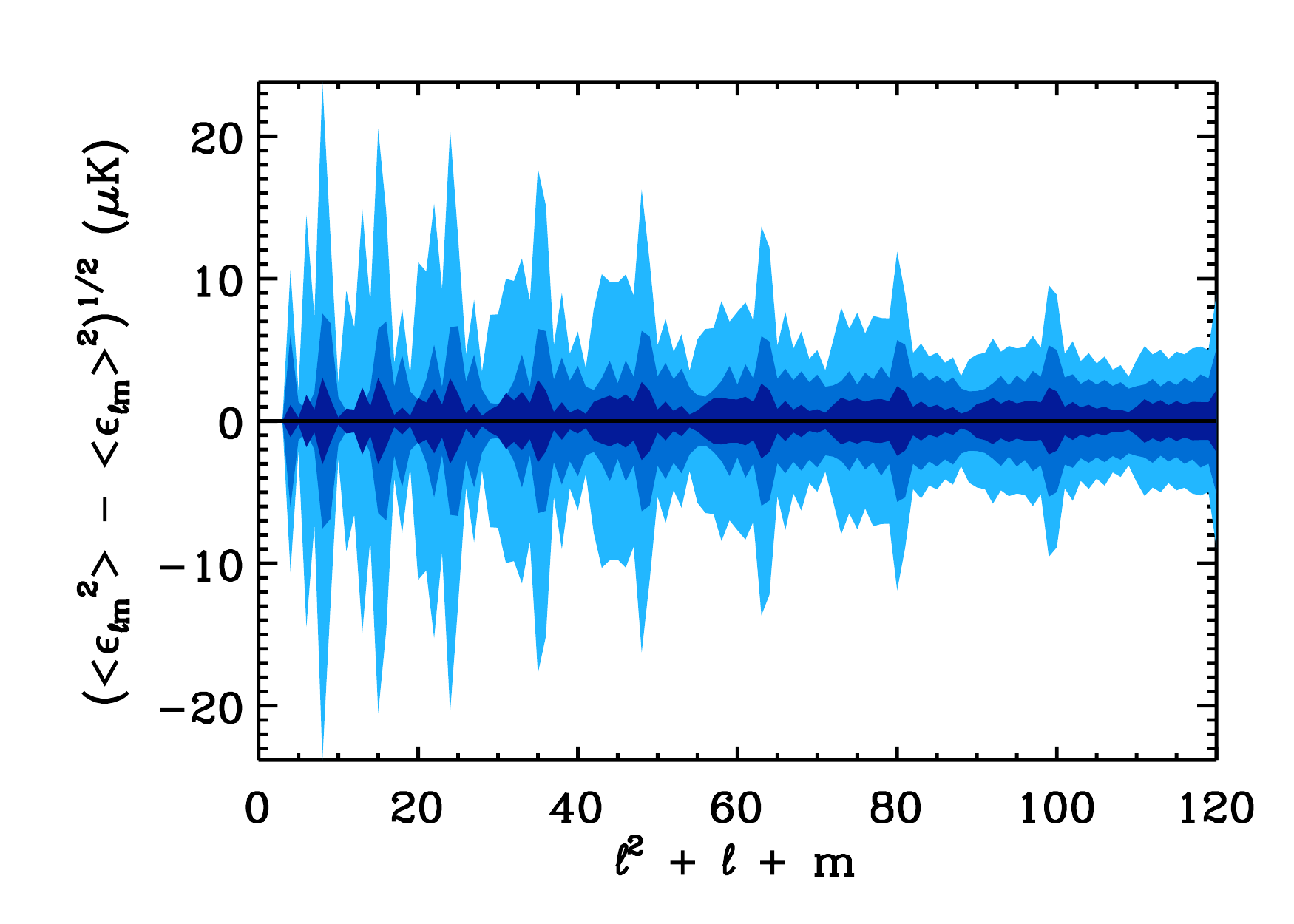}
\caption{The one-standard-deviation ranges of the reconstruction error $\epslm$ for the Gaussian ML $\almest$s (narrowest, deep-blue band), $5^{\circ}$ Top-Hat ML $\almest$s (widest, light-blue band), and Top-Hat WF $\almest$s (intermediate, mid-blue band).}
\label{fig:alm_est_err_sd}
\end{figure}

\begin{figure}[tb]
\includegraphics[width=8.5 cm]{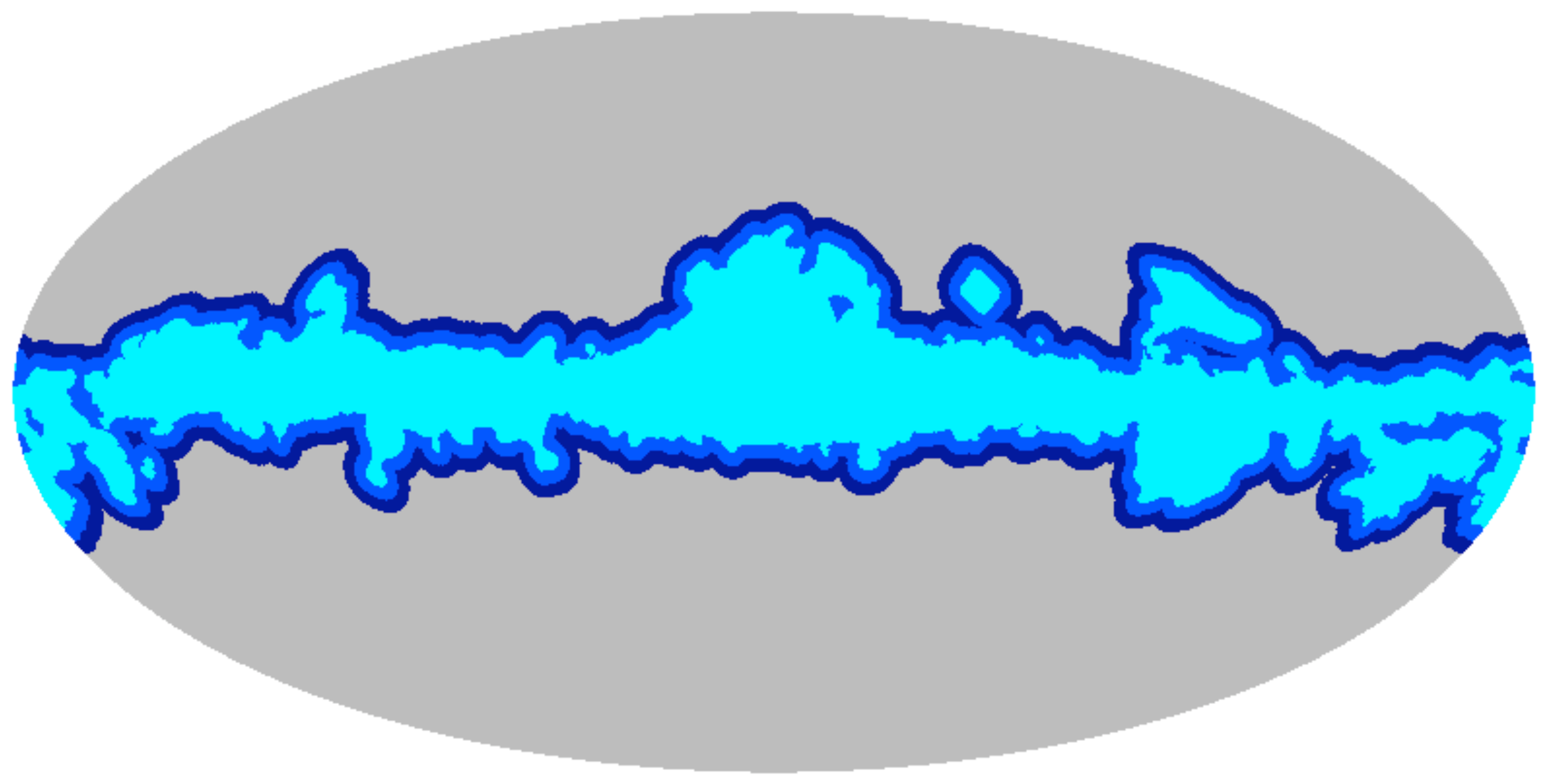}
\includegraphics[width=4.25 cm]{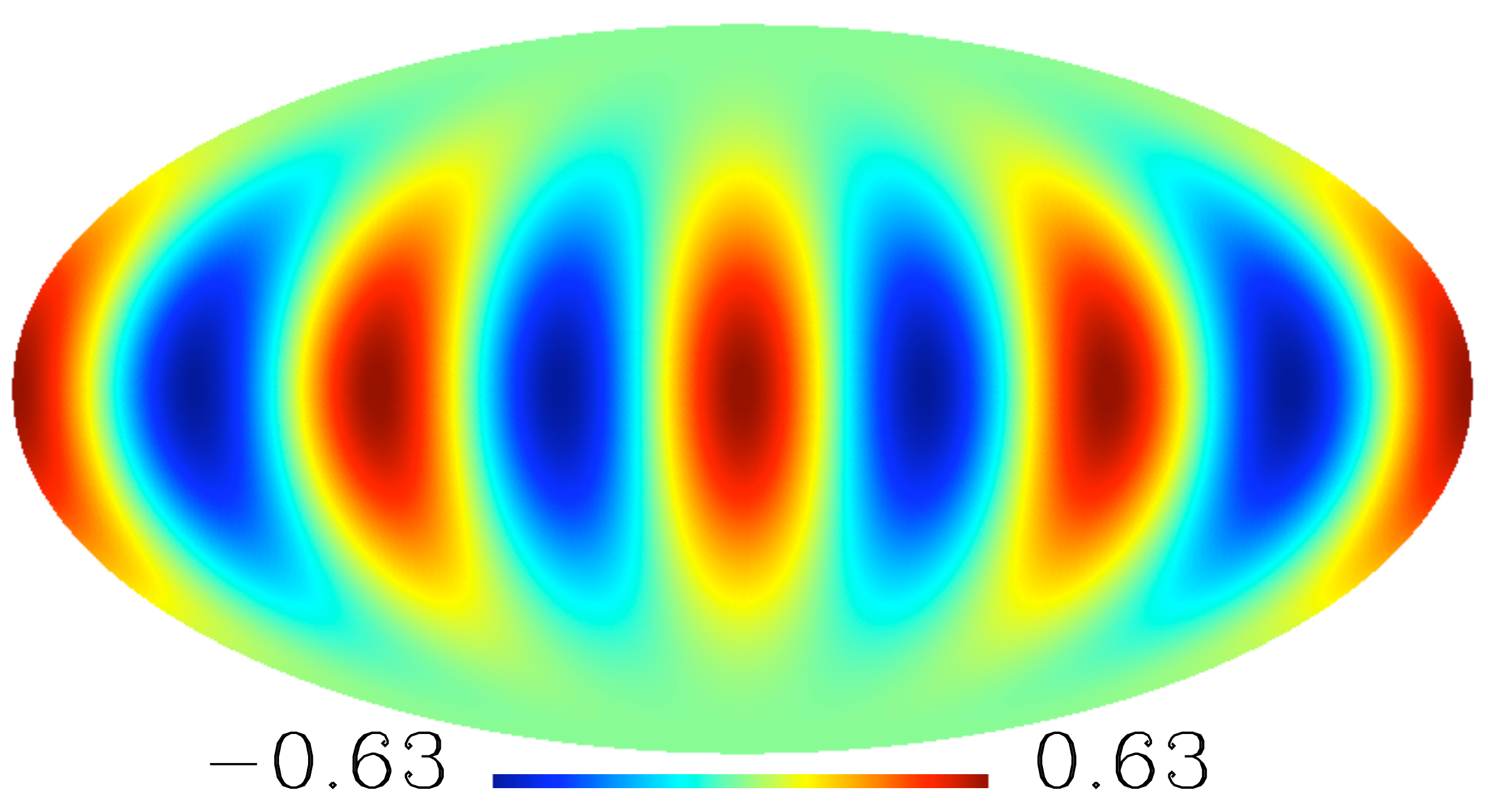}
\includegraphics[width=4.25 cm]{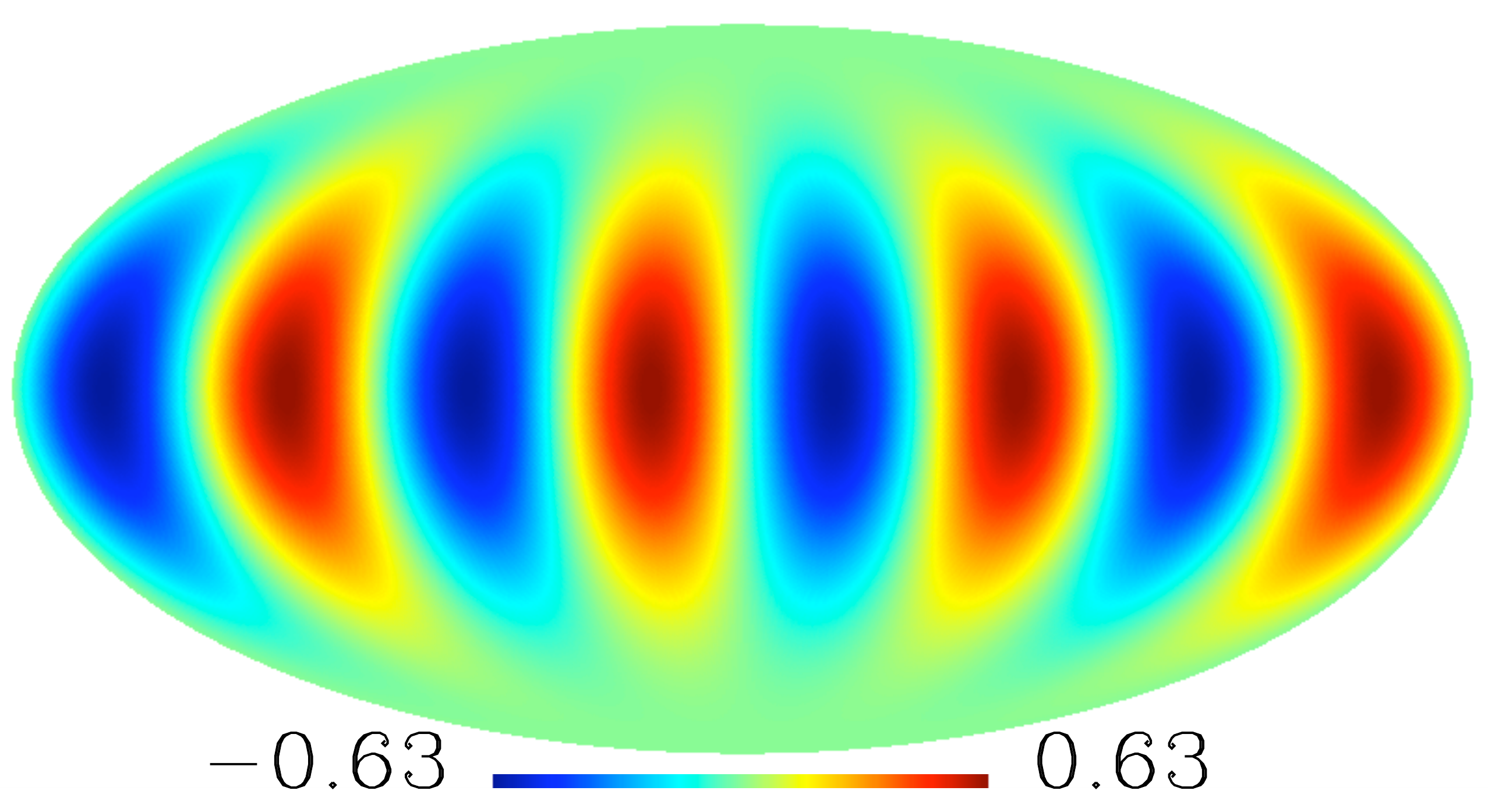}
\caption{Top: the WMAP 7-year Galaxy-only KQ85 mask (light-blue central region) extended by $2.5^{\circ}$ (mid-blue) and $5^{\circ}$ (dark-blue). Bottom: the real spherical harmonics $Y_{4\, 4}$ and $Y_{4\,-4}$. The concentration of $Y_{\ell\, \pm \ell}$ mode power towards the equator results in increased estimator variance in those modes.}
\label{fig:mask_ylm}
\end{figure}

Although the bias illustrated in Fig.~\ref{fig:alm_est_bias} looks problematic, its amplitude is at least reduced over using the contaminated full sky, and so the simple Gaussian-ML procedure may yet turn out to be useful. To compare with other possible approaches, we first need to discuss its standard deviation -- i.e. the scatter induced by the $\ell>\ellmaxrecon$ modes -- which is calculated using Eq.~\ref{eq:recon_err_ml_var} and plotted (as the deep-blue narrowest band) in Fig.~\ref{fig:alm_est_err_sd}. This is a few $\muk$ at most. For a given $\ell$, the modes that are reconstructed with the least precision are those with $| m | = \ell$, with the $m = \ell$ modes typically the worst. This confirms the observation in \citeauthor{Pontzen:2010uq}~(\citeyear{Pontzen:2010uq})~\cite{Pontzen:2010uq} that the sky cut removes the most information from modes with power concentrated towards the equator, and particularly those with extrema at $\phi = 0^{\circ}$, where the mask is at its widest~\cite{Copi:2011pe}. The mask is plotted for reference in Fig.~\ref{fig:mask_ylm}, along with examples of the affected modes. We see that, typically, three modes per $\ell$ will have increased bias or variance, but for most modes both the mean reconstruction error and its variance will be small.

In order to compare estimators, we must first quantify their performance over the range of multipoles considered. The performance measure 
\begin{eqnarray}
\zlm & = & \langle  \epslm^2 \rangle\nonumber \\
 & = & {\rm mean} ( \epslm )^2 + {\rm var} ( \epslm )\label{eq:z_lm}
\end{eqnarray}
provides the expected size of the reconstruction error $\epslm$ for each mode: summing over all modes
\begin{equation}
\ztot = \sum_{\ell,\,m} \zlm
\label{eq:z_lm_tot}
\end{equation}
therefore yields a complete measure of each estimator's performance. Any alternative estimator which removes the smoothing-induced bias should be preferred only if its $\ztot$ value is lower than that of the Gaussian ML reconstruction, and, indeed, the contaminated full-sky $\alm$s. In fact, using the $1\%$ ILC-V foreground residuals, $\ztot \simeq 33\, \muk^2$ using the full-sky $\alm$s, compared to $\sim 265\, \muk^2$ for the Gaussian ML estimator. If the residual foregrounds employed in this work are an accurate reflection of those present in the WMAP 7-year ILC map, then the contaminated full-sky $\alm$s provide a better estimate of the cosmological signal than the Gaussian ML reconstruction. The second form of Eq.~\ref{eq:z_lm} shows that both bias (see Fig.~\ref{fig:alm_est_bias}) and variance (see Fig.~\ref{fig:alm_est_err_sd}) in the reconstruction increase the value of $\ztot$. The variance term is independent of the contamination, while the bias scales linearly with the contamination. Therefore, if the amplitude of residual foregrounds in the ILC map is higher than in our illustrative example, the reduction in bias due to the use of the Gaussian ML $\almest$s will eventually overcome the variance introduced by the reconstruction. For residual levels $3 - 4$ times higher than those used here, the Gaussian ML reconstruction should be used instead of the contaminated full-sky $\alm$s. However, as we have seen, the Gaussian ML estimator (as implemented thus far)  does not eliminate the bias due to smoothing-related leakage of contaminants from within the masked region.

\begin{figure}[tb]
\includegraphics[width=4.25 cm]{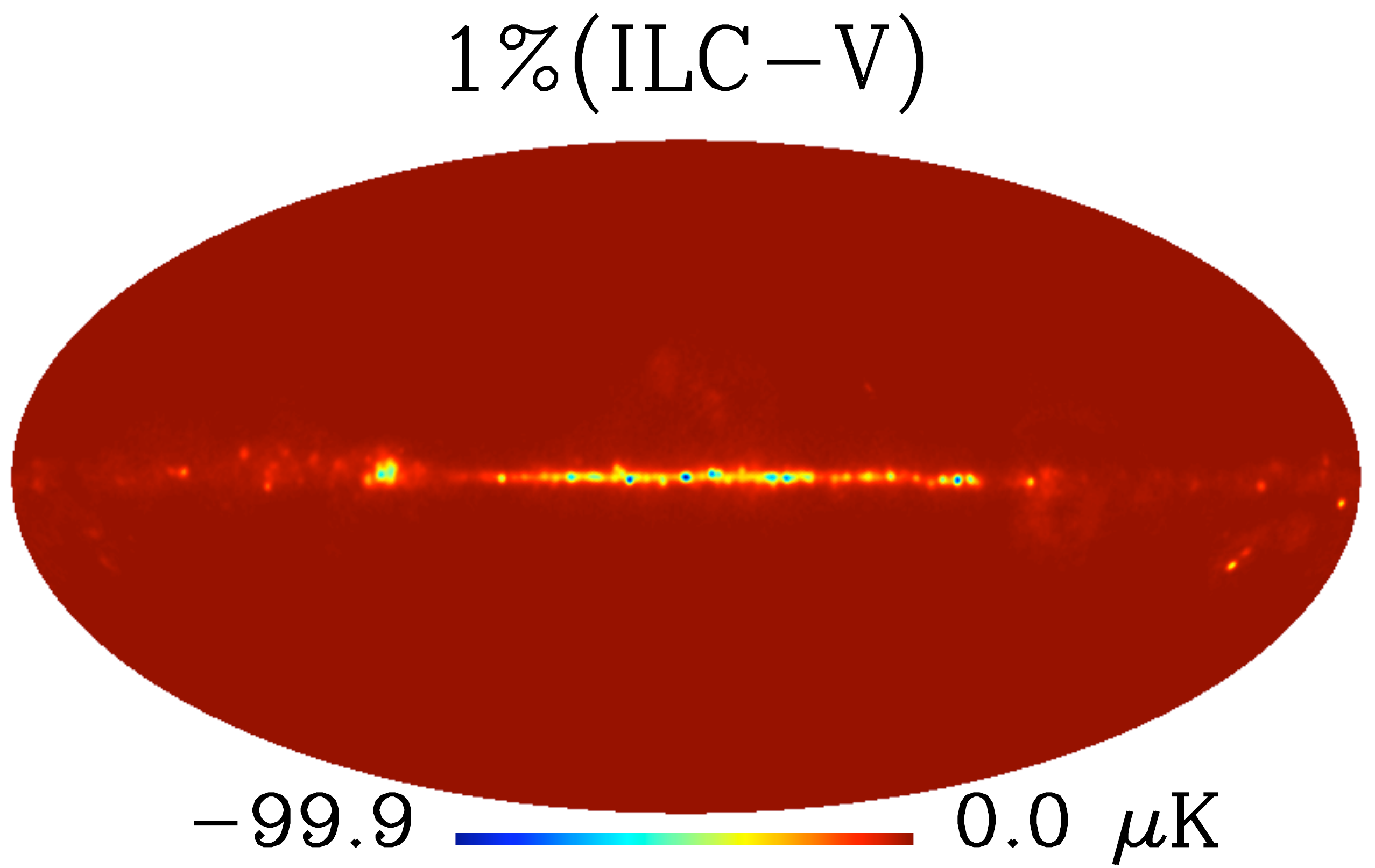}
\includegraphics[width=4.25 cm]{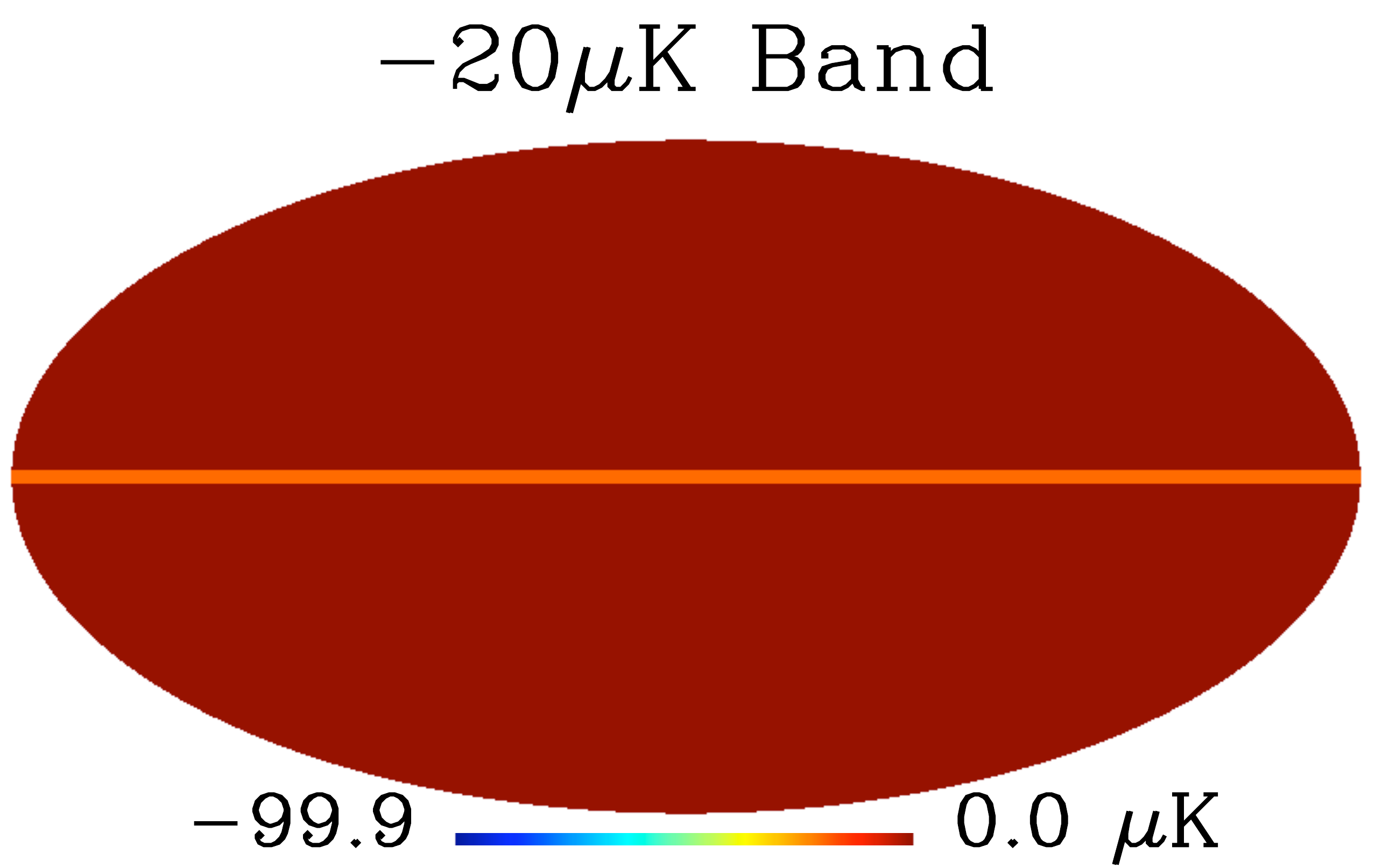}
\includegraphics[width=4.25 cm]{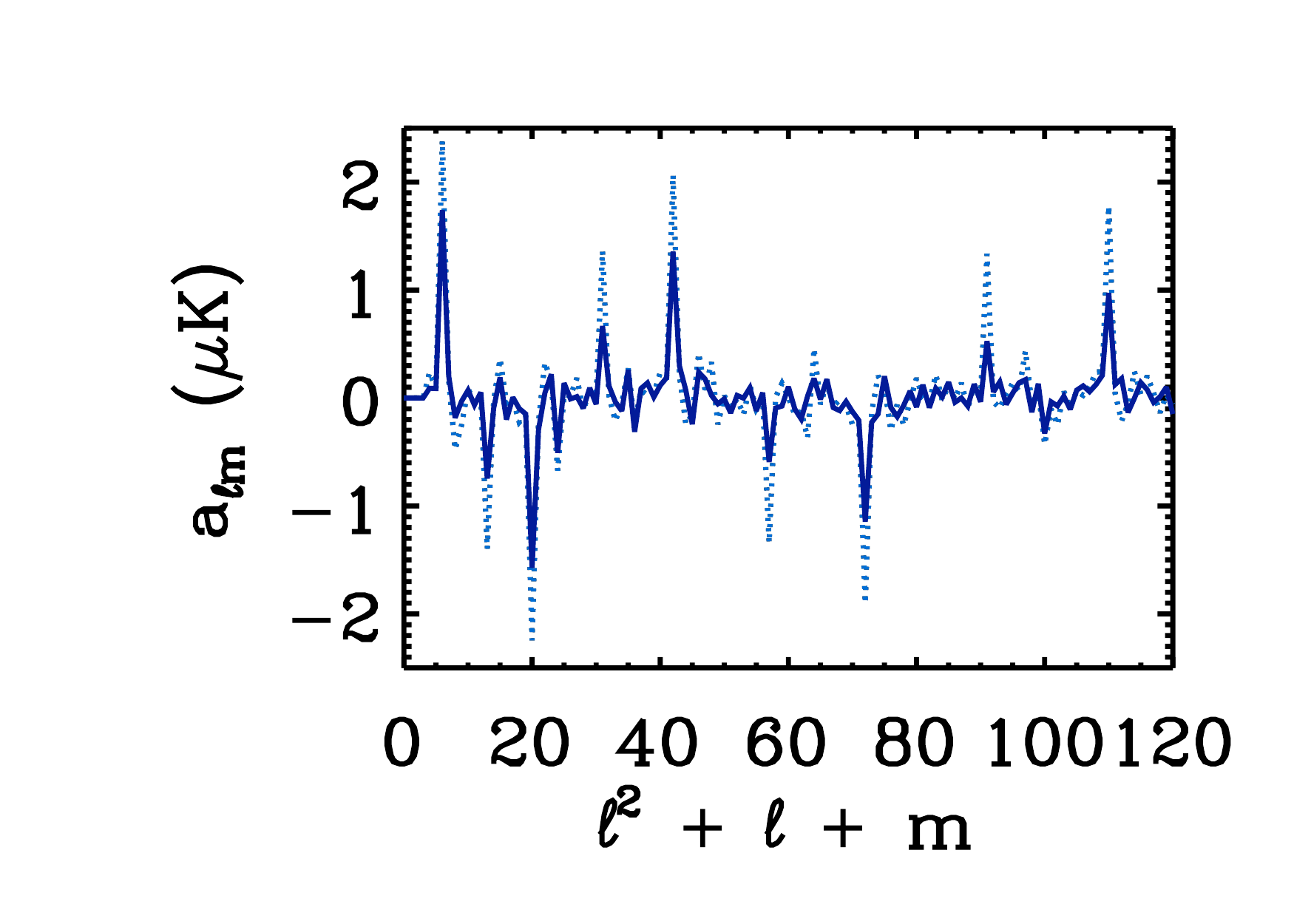}
\includegraphics[width=4.25 cm]{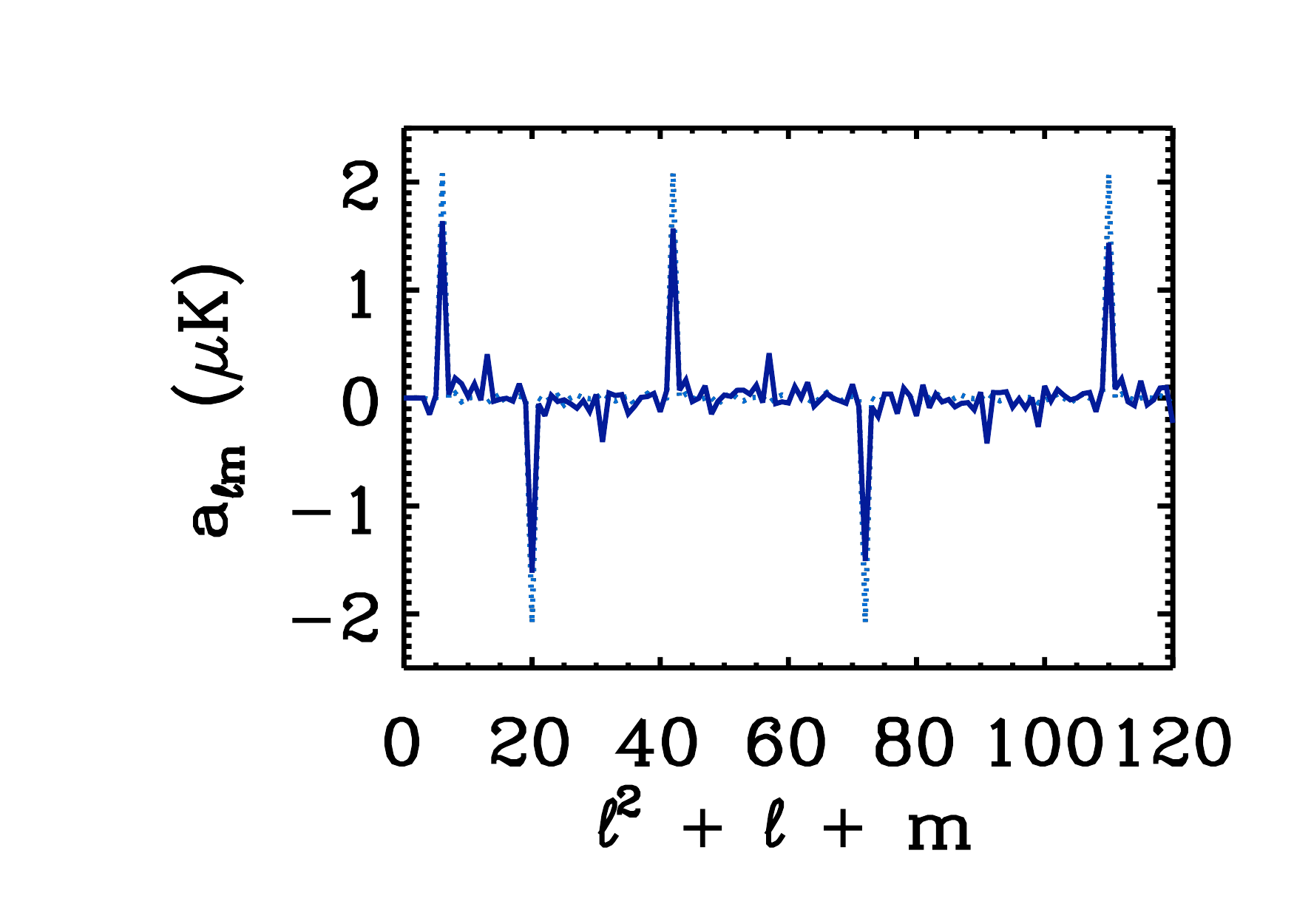}
\caption{Top: the residual foreground map employed in this work (left), and a simple model comprising a $3^{\circ}$-wide $-20\,\muk$ equatorial band. Bottom: the full-sky $\alm$s (dotted) and reconstructed Gaussian ML $\almest$s (solid) of the residual foreground maps. The simple band model captures most of the features of the smoothing-induced bias injected by the more-complex residuals.}
\label{fig:cont_comparison}
\end{figure}

While our simulated foreground residuals are simply meant to be indicative, we nevertheless expect that the smoothing bias is mainly sensitive to the {\em amplitude} of the residuals, and not their precise morphology. This can be seen in Fig.~\ref{fig:cont_comparison}, where we have modeled the residuals as a simple bar in the Galactic plane, while rescaling the amplitude to match our 1\%(ILC-V) model. This highly simplified model is able to capture most of the features of the bias in harmonic space, as seen in the lower panels of this Figure.

\section{Eliminating the bias}

At this stage, we are presented with something of a conundrum: smoothing is essential to the reconstruction process, but it is exactly this smoothing that is biasing the results. The simplest solution to this issue is to remove the areas of the sky that are within one smoothing scale of the main Galactic sky cut. However, the smoothing kernel typically used in the standard reconstruction algorithm is a Gaussian, with support across the full sky in pixel space, and the set of contaminated pixels is hence poorly defined. This problem can be solved by using a kernel with finite pixel-space support -- for simplicity we choose a top-hat -- as all contaminated pixels fall within a kernel's radius of the mask.

Na\"ively selecting the diameter of the top-hat smoothing kernel to be $10^{\circ}$, we reconstruct the simulated foreground residuals using the Galaxy-only KQ85 mask extended by $5^{\circ}$ -- hereafter the ``$10^{\circ}$ Top-Hat ML'' reconstruction. The results are plotted in light blue (pale line along $x$-axis) in Fig.~\ref{fig:alm_est_bias}: the smoothing-induced bias has been eliminated. However, the measure of reconstruction quality has deteriorated dramatically to $\ztot \simeq 11\,252\,\muk^2$, significantly worse than the Gaussian ML reconstruction. There are two reasons for this increase in $Z$, which is now sourced entirely by increased variance in the reconstruction. Firstly, the top-hat smoothing kernel has support over a greater range of multipoles than the Gaussian kernel, and so more ambiguous modes contribute to the covariance matrix (in this case, and for all further kernels, we increase $\ellmax$ to $1024$ to capture all relevant modes, even though at $10^{\circ}$ the smoothing kernel is effectively band-limited at $\ell \sim 100$).\footnote{Truncating the kernel by selecting a lower $\ellmax$ leads to $\order(10\%)$ variations in the results.} The second factor is that the reconstruction-error variance increases very rapidly with the area of the sky that is masked~\cite{deOliveiraCosta:2006zj}. This suggests the use of narrower smoothing kernels, although this necessarily increases the power of the high-$\ell$ noise.

\begin{figure}[tb]
\includegraphics[width=8.5 cm]{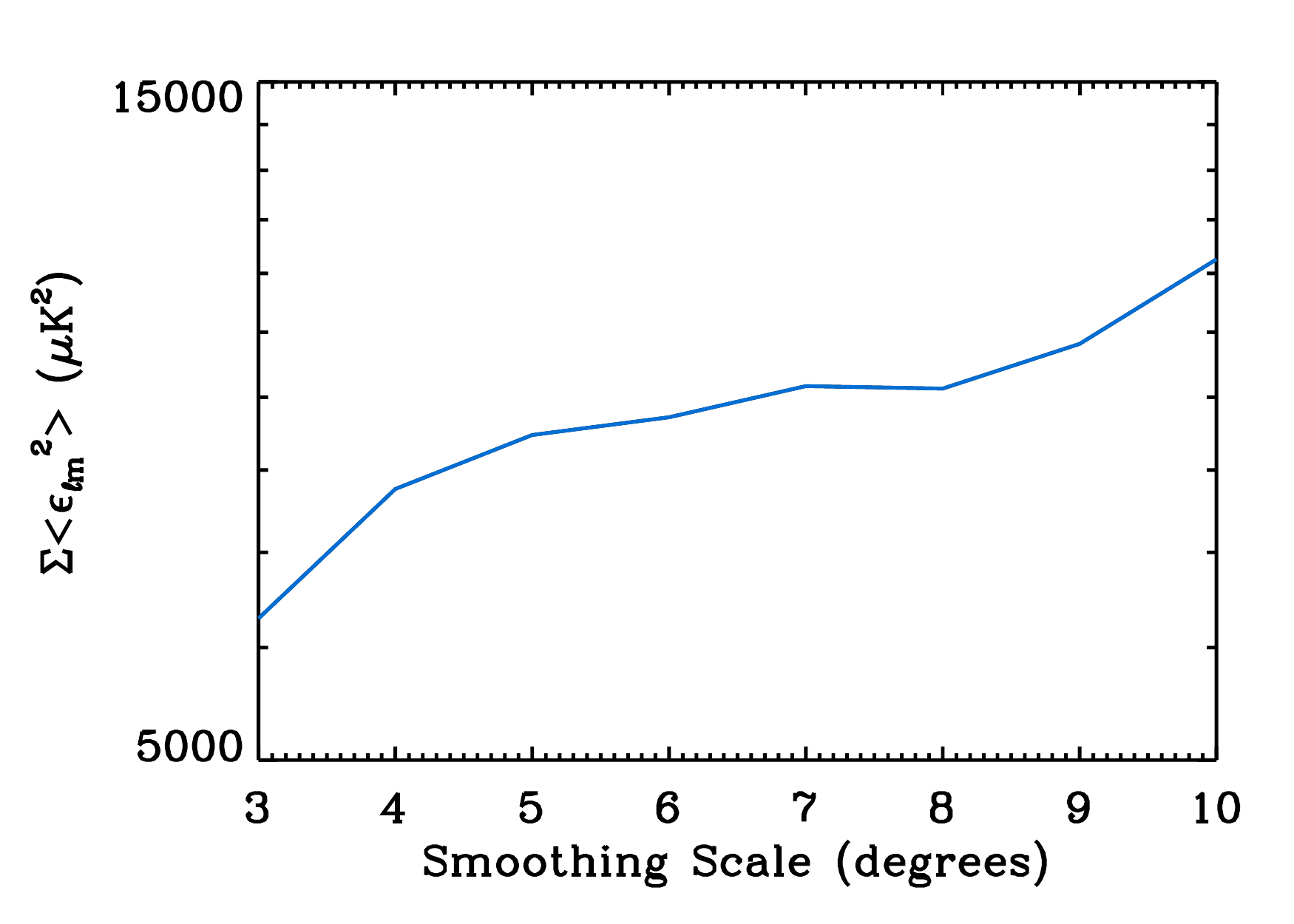}
\caption{The impact of mask extension and smoothing-kernel diameter on the quality of the bias-free maximum-likelihood reconstruction. Reconstruction-error variances are calculated using top-hat smoothing kernels of varying diameter, and extending the KQ85 mask by one kernel radius each time. Smoothing at the lowest-possible scale will produce the most faithful reconstruction. Note that the curve is not smooth as both the extent and {\em shape} of the mask change as it is extended.}
\label{fig:smooth_ext_z_dep}
\end{figure}

The interplay between the variance injected by decreased smoothing and increased masking is shown in Fig.~\ref{fig:smooth_ext_z_dep}. Here, the total reconstruction-error variance is plotted for top-hat smoothing kernels of diameter $3^{\circ} - 10^{\circ}$, and hence mask extensions of $1.5^{\circ} - 5^{\circ}$. The reduction in variance due to minimizing the sky cut dominates the added noise from narrower smoothing. We therefore select the width of the smoothing kernel to be as small as possible, given the resolution of the degraded map. At $\nside = 16$, the pixels are $\sim 3^{\circ}$ across, so to avoid injecting bias through pixelization~\cite{Copi:2011pe} (which would not be captured by Eq.~\ref{eq:recon_err_ml_var}) we choose our optimal kernel diameter to be just larger: $5^{\circ}$. Hereafter, we refer to this reconstruction -- using the $5^{\circ}$ top-hat smoothing kernel and Galaxy-only KQ85 mask extended by $2.5^{\circ}$ -- as the ``$5^{\circ}$ Top-Hat ML'' reconstruction.

The reconstruction-error variance calculated for the $5^{\circ}$ Top-Hat reconstruction is plotted as the light-blue outermost region of Fig.~\ref{fig:alm_est_err_sd}. Even using the minimum possible mask extension the reconstruction-error variance is still an order of magnitude larger than that of the Gaussian ML estimator. In terms of the measure of reconstruction quality, the biased Gaussian ML reconstruction ($\ztot \sim 265\, \muk^2$) should be strongly preferred to the $5^{\circ}$ Top-Hat ML case ($\ztot \sim 8466 \,\muk^2$) for residual levels comparable to those used in this work. As the Top-Hat ML reconstruction is unbiased, this value of $\ztot$ is fixed (for a given sky cut). Thus, only if the residuals are greater than $\sim 25\%$ ILC$-$V will the $5^{\circ}$ Top-Hat ML reconstruction outperform the Gaussian ML reconstruction. Note that the quality of the reconstruction could be improved further if it was performed at higher resolution, as smaller smoothing kernels could be used. This will necessarily have to be traded off against the increased computational requirements.

\section{Reducing the variance}

The increase in variance encountered when using extended sky cuts is far beyond that expected due to the reduction in pixel count: reducing $\fsky$ from $81\%$ to $74\%$ should, assuming uncorrelated pixels for simplicity, increase the variance by only $\sim 10\%$. The dominant issue is that the maximum-likelihood reconstruction allows the temperature field in the masked region of the sky to have infinite variance. For small sky cuts (and small $\ellmaxrecon$) this is fine: one cannot ``hide'' large-scale power within the cut, and so the variance on the large-scale $\almest$s is low. Extending the KQ85 mask not only increases its overall width, but also closes a number of small gaps that allow the estimator limited access to the poorly-constrained equatorial modes (see Fig.~\ref{fig:mask_ylm}). The estimator is therefore free to fill the cut with significant low-$\ell$ power (compare Fig.~\ref{fig:ilc_recon} top-right and bottom-left), and the estimator variance rises rapidly.

The variance of the reconstruction error can be reduced by enforcing a prior on the power within the Galactic cut using Wiener-filtering (Method 5 in Ref.~\cite{Tegmark:1996qs}; see also Refs.~\cite{Bielewicz:2004en, Wiener:1964, Bunn:1994xn, Zaroubi:1994mx}). The Wiener-filtered $\almest$s then maximize the posterior probability of reconstructing the underlying $\alm$s, given the trusted data and theoretical power spectrum. In practice, this is achieved by adding a theoretical (inverse) covariance matrix for the multipoles of interest to the reconstruction matrix
\begin{equation}
\label{eq:wf_recon_mat}
\wmat = [ \smat^{-1} + \ymat^t \cmat^{-1} \ymat ]^{-1} \ymat^t \cmat^{-1}.
\end{equation}
$\smat$ here is a diagonal $\nalm \times \nalm$ matrix with elements equal to the smoothed theory power spectrum $\clsmooth$. The theory prior restricts this ``Top-Hat WF'' reconstruction from filling the sky cut with arbitrary power (see Fig.~\ref{fig:ilc_recon} bottom-right), and reduces the variance of the estimator's reconstruction error accordingly (see the mid-blue intermediate region in Fig.~\ref{fig:alm_est_err_sd}). The analytic expectation for the reconstruction-error variance is
\begin{equation}
\label{eq:recon_err_wf_var}
\langle \epslmvec \epslmvec^t \rangle - \\
\langle \epslmvec \rangle \langle \epslmvec^{t} \rangle = \\
\smat \, [ \smat + \nmat ]^{-1} \nmat,
\end{equation}
where $\nmat$ is the noise power spectrum, defined to be $\nmat = [ \ymat^{t} \cmat^{-1} \ymat ]^{-1}$ (\emph{i.e.} the variance of the Top-Hat ML reconstruction error). By adding a theory prior to the power within the sky cut, and hence requiring finite power in that region, the Wiener-filtered reconstruction tends to produce $\almest$s that are closer to zero than the maximum-likelihood case. While this could be seen as biasing the $\almest$s toward lower values\footnote{Assuming for clarity zero noise, the Wiener-filtered reconstruction yields $\almestvec = \smat [ \smat + \nmat ]^{-1} \almvec$, {\em i.e.} a multiplicative bias. Note that the ensemble average $\langle \almestvec \rangle = \langle \almvec \rangle = 0$.}, it can also be interpreted as being conservative, and applying the prior belief that the information within the mask is similar to the trusted information outside the mask. In other words, we should be happy to trade off a small multiplicative bias against a significant reduction in variance.

This is automatically encapsulated in the measure of reconstruction quality $\ztot$ for the $5^{\circ}$ Top-Hat WF estimator, which has improved to $\sim 1521 \,\muk^2$. However this is still worse than that of the Gaussian ML estimator. The bias arising in the Top-Hat WF reconstruction is not from smoothing but from a prior, so $\ztot$ is fixed for a given mask, and always lower than that of the corresponding Top-Hat ML reconstruction. For contamination levels of $\gtrsim 10\%$ ILC$-$V (such as those found in the foreground-reduced maps for the individual WMAP frequency bands), the $5^{\circ}$ Top-Hat WF reconstruction therefore represents the most reliable estimator considered in this work.

\begin{figure}[tb]
\includegraphics[width=8.5 cm]{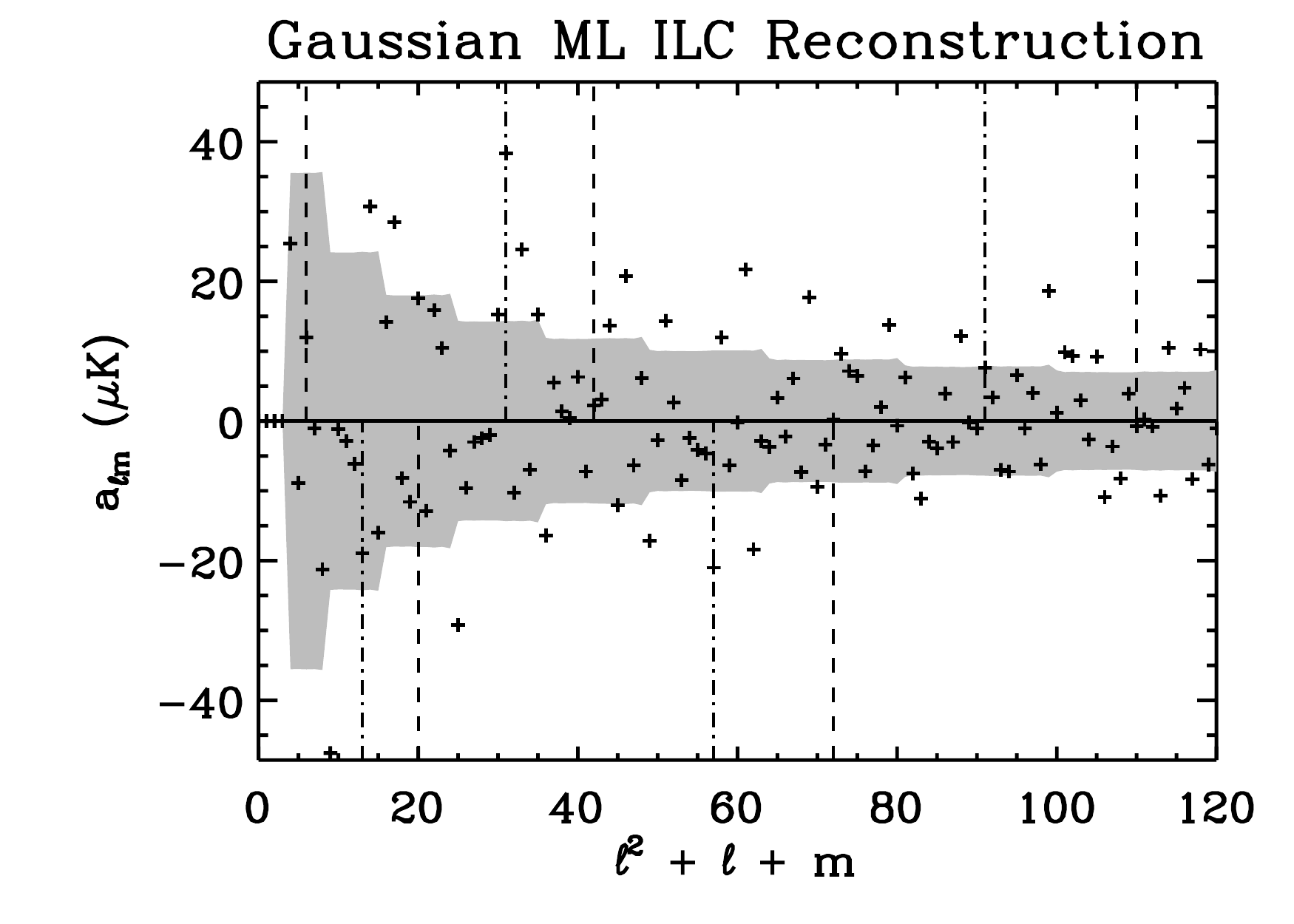}
\includegraphics[width=8.5 cm]{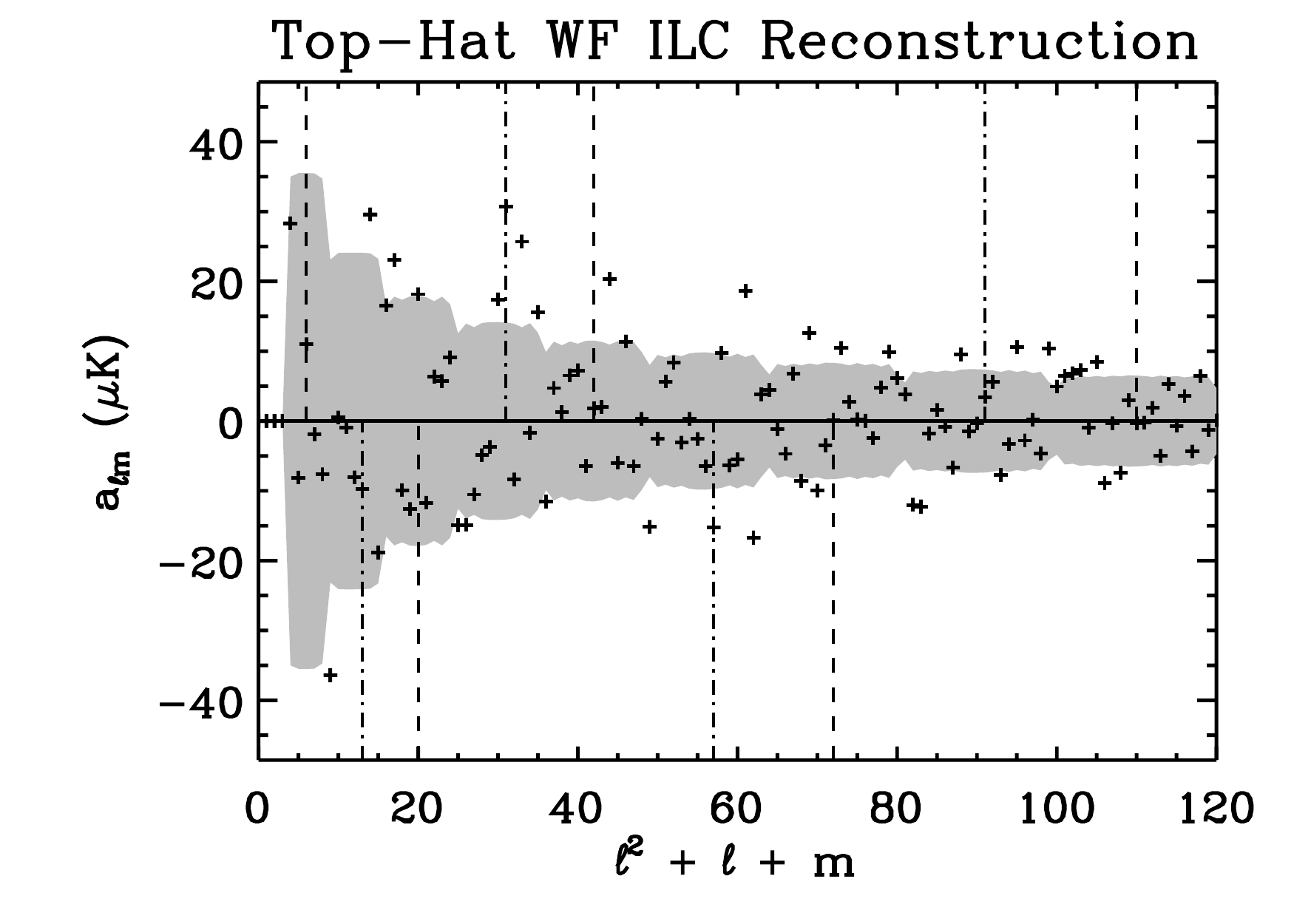}
\caption{The reconstructed WMAP 7-year ILC $\almest$s, calculated using the Gaussian ML reconstruction (top) and the $5^{\circ}$ Top-Hat WF reconstruction (bottom). The shaded areas represent the estimator standard deviations. The modes that are most contaminated by the simulated foregrounds in the Gaussian ML reconstruction are indicated, along with their expected sign, by dashed ($Y_{{\rm even}\,0}$) and dash-dotted ($Y_{{\rm odd}\,1}$) lines.}
\label{fig:ilc_alm_plus_std}
\end{figure}

We do not know the precise magnitude or morphology of the residual foregrounds in the WMAP data. We can, however, examine the Gaussian ML and Top-Hat WF reconstructions of the ILC by eye to determine if there is an obvious impact due to residual foregrounds. The $\almest$s generated from these two reconstructions are plotted in Fig.~\ref{fig:ilc_alm_plus_std}, along with the {\em estimator} standard deviation ($\sqrt{\langle | \almest |^2 \rangle - \langle \almest \rangle^2}$). Comparing the two plots, we see that there is very little difference between the $\almest$s returned in each case. Further, no modes look statistically anomalous at the $3\sigma$-level, even those that we expect to be contaminated from the simple residuals model used here.

\section{Relation of $\bf{\alm}$ reconstruction to the QML estimator for the $\bf{\cl}$s}

We have so far discussed estimating the full-sky $\alm$s from cut-sky data, which is equivalent to reconstructing the smoothed temperature field. However, in the context under which the smoothing-induced bias was revealed~\cite{Efstathiou:2009di, Aurich:2010gw,Copi:2011pe} it is in fact only the angular power spectrum $C_{\ell}$ of the temperature field which is required.

A popular method for estimating the full-sky angular power spectrum is to adopt the quadratic maximum-likelihood estimator as first derived in Ref.~\cite{Tegmark:1996qt}. It has been noted (see e.g. Section 3 of Ref.~\cite{Efstathiou:2009di} for a complete discussion) that the QML estimator can be formed using the maximum-likelihood $\almest$s. On the surface, the QML estimates (henceforth denoted $\clqml$) may therefore seem to be susceptible to similarly problematic contamination from a smoothing stage.

However, this is not the case: in fact the $\clqml$s are far more robust to the content of the cut because the smoothing can be conducted on vastly smaller scales (e.g. 1$^\circ$ in \citeauthor{Pontzen:2010uq} (\citeyear{Pontzen:2010uq}) \cite{Pontzen:2010uq}). Note that \citeauthor{Copi:2011pe} (\citeyear{Copi:2011pe}) \cite{Copi:2011pe} miss this point, because they consider only two extreme cases: (i) smoothing at 10$^\circ$ and (ii) failing to smooth. They therefore reach the erroneous conclusion that the QML estimator is susceptible to contamination from within the mask. We explicitly verified that the pipeline used by \citeauthor{Pontzen:2010uq} (\citeyear{Pontzen:2010uq}) \cite{Pontzen:2010uq} is independent of any contamination placed fully inside the mask.

The above paragraphs at first appear to be contradictory, since they simultaneously claim (a) that the QML power spectrum estimates can be formed out of the ML temperature field reconstruction; and (b) that the QML power spectrum estimates can still be constructed from maps smoothed on degree scales (whereas the $\almest$s will necessarily become noisy for sufficiently high $\ellmaxrecon$). However, this is not a true contradiction because the QML estimates are not formed directly from the noisy $\almest$s, but rather through an expression (Eq. 23 of Ref.~\cite{Efstathiou:2009di}) which specifically downweights poorly constrained modes. It is this cautious treatment of ambiguous modes which makes power spectrum estimation, as opposed to $\alm$ reconstruction, so well-behaved, irrespective of the shape of the smoothing kernel employed.

\section{Discussion}

Maximum-likelihood estimators, $\almest$, are often used to reconstruct the large-scale spherical harmonic coefficients, $\alm$,  from partial-sky data. The technique relies on smoothing to restrict the amount of small-scale noise accessible to the reconstruction, but smoothing has been shown to contaminate ``clean'' pixels with residual foregrounds from within the sky cut. In this work, we have examined the impact of this smoothing-induced bias on the maximum-likelihood reconstruction. We have shown that it is possible to mitigate the bias by removing the contaminated regions, but these are only well-defined if smoothing is performed using a kernel with finite support on the sky. This precludes the use of the commonly used Gaussian kernel.  Cutting a larger portion of the sky greatly increases the variance of the reconstruction, but it is possible to counteract this effect by enforcing a prior on the reconstructed coefficients using a Wiener filter. We have therefore proposed an estimator -- using top-hat smoothing, extended masks and a Wiener-filtered reconstruction -- which does not suffer from smoothing-induced bias. By considering the expectation of the square of the reconstruction error, $\ztot = \sum_{\ell,\,m} \langle (\almest - \alm)^2 \rangle$, we have compared the performance of the maximum-likelihood and Wiener-filtered estimators in the presence of simulated CMB foreground residuals.

The reconstruction performance measure $\ztot$ scales with the estimators' bias and variance, which in turn are governed by the amplitude of contamination and the size of the sky cut, respectively. The fiducial maximum-likelihood reconstruction is performed using relatively small sky cuts, but is susceptible to contamination through smoothing-induced bias; the finite-smoothing Wiener-filtered reconstruction does not suffer from smoothing-induced bias, but makes use of extended masks. Increasing the level of contamination therefore increases $\ztot$ for the maximum-likelihood reconstruction only, which suggests that there is a level of contamination above which one should switch from the maximum-likelihood to the Wiener-filtered reconstruction.

Given an estimate of the morphology and amplitude of the contaminants within the cut sky, one can predict which modes will be biased and by how much, and hence determine the threshold at which one should swap estimators. We find that this threshold is relatively insensitive to the precise morphology of foreground residuals at large scales, and is mainly governed by their amplitude. Calculating $\ztot$ for the two estimators in the presence of estimated foreground residuals, we determine this threshold to be $\sim 10$ times the amplitude of the foreground residuals used in this work. Assuming that the ILC contains similar levels of contamination to those used here, we therefore recommend the use of either the contaminated full-sky $\alm$s or the fiducial maximum-likelihood $\almest$s when handling this data-set. However, when using foreground-reduced maps for individual WMAP frequencies, which contain much greater foreground residuals, the Wiener-filtered reconstruction will provide the best estimate of the large-scale underlying CMB signal. Note that, as the Wiener-filtered $\almest$s are a maximum-{\em posterior} solution, care must be taken if the reconstruction output is being used for further model-selection steps. The reconstruction techniques are, however, most commonly used to test the null hypothesis, in which case the prior employed in this work is completely appropriate.

For problems requiring only a power spectrum (as opposed to the full temperature field) the issues described in this paper are essentially irrelevant because the smoothing can be conducted on vastly smaller scales, the resulting range of poorly constrained modes being automatically downweighted.

\acknowledgements
We thank George Efstathiou for interesting discussions. SMF is supported by the Perren Fund and STFC. HVP is supported by Marie Curie grant MIRG-CT-2007-203314 from the European Commission, and by STFC and the Leverhulme Trust. AP is supported by Emmanuel College, Cambridge. We acknowledge use of the HEALPix package and the Legacy Archive for  Microwave Background Data Analysis (LAMBDA).  Support for LAMBDA is provided by the NASA Office of Space Science.

\bibliography{paper}

\end{document}